\documentclass[aps, prb, twocolumn, amssymb, amsmath, showpacs, superscriptaddress]{revtex4-1}
\usepackage{bm}
\usepackage{times}
\usepackage{graphicx}
\usepackage{makecell}
\usepackage{color}
\usepackage{dcolumn}
\usepackage[colorlinks=true, letterpaper=true, pdfstartview=FitV, linkcolor=blue, citecolor=blue, urlcolor=blue]{hyperref}
\usepackage{appendix}
\usepackage[normalem]{ulem}

\newcommand{\vk}{\nabla_{\!k}}
\newcommand{\bd}{\hat{\bm d}}
\newcommand{\vect}[1]{\boldsymbol{#1}}
\newcommand{\dd}{\mathrm{d}}
\newcommand{\Tr}{\mathrm{Tr}}

\newcommand{\eps}{\varepsilon}

\usepackage{helvet}

\usepackage{pifont}

\begin{document}

\title{Curvature-Flux Representation of Local Dirac-Node Topology from Non-Hermitian Zeeman Quantum Geometry}
\author{Rongjie Cui}
\thanks{These authors contributed equally to this work.}
\affiliation{College of Physics and Optoelectronic Engineering, Shenzhen University, Shenzhen 518060, China}
\author{Longjun Xiang}
\thanks{These authors contributed equally to this work.}
\affiliation{College of Physics and Optoelectronic Engineering, Shenzhen University, Shenzhen 518060, China}
\author{Fuming Xu}
\affiliation{College of Physics and Optoelectronic Engineering, Shenzhen University, Shenzhen 518060, China}
\affiliation{Quantum Science Center of Guangdong-Hongkong-Macao Greater Bay Area, Shenzhen 518045, China}
\author{Jian Wang}
\email{jianwang@hku.hk}
\affiliation{College of Physics and Optoelectronic Engineering, Shenzhen University, Shenzhen 518060, China}
\affiliation{Quantum Science Center of Guangdong-Hongkong-Macao Greater Bay Area, Shenzhen 518045, China}
\affiliation{Department of Physics, The University of Hong Kong, Pokfulam Road, Hong Kong, China}

\begin{abstract}
Local Dirac-node topology is usually described by a winding number or Berry phase, whereas curvature-flux formulations are commonly associated with global band topology. Here we show that the non-Hermitian Zeeman quantum geometric tensor (QGT) provides a curvature-flux representation of local $\pi_1$ topology. Unlike the conventional Hermitian QGT, whose metric and Berry curvature are fixed respectively by its real symmetric and imaginary antisymmetric parts, the Zeeman QGT must be decomposed according to symmetry under index exchange. This decomposition yields normal and anomalous metric-curvature sectors. The normal sector has the same metric-curvature structure as the conventional Hermitian QGT. The anomalous sector, by contrast, contains two components absent in the conventional QGT: an imaginary symmetric metric-like tensor and a real antisymmetric curvature-like tensor. In a two-dimensional Dirac system, the anomalous Zeeman curvature forms a radial flux field Hodge-dual to the tangential winding field of the Dirac node, thereby converting the local winding invariant into a Gauss-type flux invariant. We further show that the four Zeeman-geometric sectors map one-to-one onto frequency- and symmetry-resolved gyrotropic conductivity channels, with reciprocal kinetic magnetoelectric response providing a complementary probe. These results establish non-Hermitian Zeeman quantum geometry as a measurable framework connecting local Dirac-node topology and transport.
\end{abstract}

\maketitle

\section{Introduction.}
The band topology of quantum materials\cite{Kane2010,SCZhang2011,Das2016} can be understood from both local and global perspectives. Near an isolated band crossing, the relevant physics is encoded in geometric quantities defined in a restricted region of momentum space. For example, a two-dimensional Dirac node can be characterized by a quantized Berry phase or winding number evaluated on a closed loop enclosing the singularity, thereby realizing a $\pi_1$-type topological characterization\cite{Mermin1979,Nakahara}. By contrast, global band topology emerges only after integrating geometric quantities over the full Brillouin zone. A paradigmatic example is the Chern number, obtained from the Brillouin-zone integral of the Berry curvature, which diagnoses the topology of an entire Bloch band and is commonly associated with $\pi_2$-type topology\cite{Mermin1979,Nakahara,Volovik2003}. An important open question is whether a local winding-type topology can also be recast into a curvature-flux form, so that local and global topological descriptions can be discussed within a more unified geometric language.

This question naturally points to the framework of quantum geometry. In the conventional Hermitian quantum geometric tensor (QGT), the real symmetric part defines the quantum metric, while the imaginary antisymmetric part defines the Berry curvature\cite{Xiao2010}. Together they govern adiabatic phases, orbital and electric responses, geometric constraints in Bloch bands\cite{AhnNP, Onishi2024, Moore2010, Ma2021, Liu2024, Komissarov2024, Xiang2024PRB, Tokura2018, Holder2020}, as well as nonlinear Hall effects\cite{Sodemann2015, Ma2019, Kang2019, Du2021, Lai2021, Wei2022PRB, Wei2023PRL, Zhang2023, Gao2023, Wang2023, Kaplan2024, Han2024, Jia2024, Wei2025PRL}. However, when the coupling operator is not the position operator, the corresponding QGT generally becomes non-Hermitian. For instance, this occurs in Zeeman coupling (${\bf B} \cdot {\boldsymbol\sigma}$), which involves spin matrix elements rather than electric-dipole coupling (${\bf E} \cdot {\bf r}$). Its geometric content is then no longer exhausted by a single metric-curvature pair. A notable example is the Zeeman QGT\cite{XiangZeeman,Zeeman1,Zeeman2,Zeeman3,Zeeman4}, which has recently been shown to underlie gyrotropic magnetic currents\cite{Zhong2016}.

The non-Hermitian character of this tensor reveals a geometric structure beyond the conventional Hermitian QGT. When classified by symmetry under index exchange, the Zeeman QGT separates into normal and anomalous metric-curvature sectors. The normal sector is continuously connected to the standard quantum metric and Berry curvature, whereas the anomalous sector contains a real antisymmetric curvature-like component with no counterpart in the conventional QGT [Fig.~\ref{Fig1}]. We show that, in a two-dimensional Dirac system, this anomalous Zeeman curvature carries a quantized radial flux singularity that is Hodge-dual to the tangential winding field of the Dirac node. This gives a curvature-flux representation of the same local $\pi_1$ topology that is usually described by a winding number, as illustrated in Fig.~\ref{Fig2}(a). We further show that the four Zeeman-geometric sectors can be isolated through the tensor structure and low-frequency scaling of the gyrotropic conductivity, as summarized in Fig.~~\ref{Fig2}(b),
and that the reciprocal kinetic magnetoelectric response provides a complementary probe. The same construction also extends to the non-Hermitian spin QGT.


\section{Generalized Zeeman quantum geometry}
In general, the quantum geometric tensor (QGT) for a band pair $(n,m)$ is defined as~\cite{AhnNP}
\begin{align}
T^{ab}_{nm}=r^a_{nm}r^b_{mn},
\label{QGT}
\end{align}
where $a,b\in\{x,y,z\}$, $r^a_{nm}\equiv\langle u_n|\hat r^a|u_m\rangle$ with
$\hat r^a=i\,\partial/\partial k_a$, and $|u_m\rangle$ is the cell-periodic Bloch state.
Eq.~\eqref{QGT} is gauge-invariant and generally complex. Writing
$T^{ab}_{nm}=g^{ab}_{nm}-\tfrac{i}{2}\Omega^{ab}_{nm}$ yields the
\textit{quantum metric} (QM) $g^{ab}_{nm}\equiv\mathrm{Re}[r^a_{nm}r^b_{mn}]$ and the
\textit{Berry curvature} (BC) $\Omega^{ab}_{nm}\equiv -2\,\mathrm{Im}[r^a_{nm}r^b_{mn}]$.
A gauge-invariant \textit{Zeeman QGT} was recently proposed~\cite{XiangZeeman},
\begin{align}
 T^{Z,ab}_{nm}=r^a_{nm}\,\sigma^b_{mn},
\label{ZeemanQGT}
\end{align}
where $\sigma^b_{mn}\equiv\langle u_m|\hat\sigma^b|u_n\rangle$ involves the Pauli operator $\hat\sigma^b$ along $b$. This construction probes quantum
distances in the spinor Bloch manifold and underlies gyrotropic magnetic
responses~\cite{XiangZeeman}.
Analogously, the Zeeman Berry curvature (ZBC) $-2\text{Im}[{T}^{Z,ab}_{nm}]$
and the Zeeman quantum metric $ \text{Re}[{T}^{Z,ab}_{nm}]$ were defined,
both of which play a crucial role in driving the gyrotropic magnetic current~\cite{Zhong2016,XiangZeeman}. However, as discussed below, for a rank-two quantum geometric tensor, the distinction between metric- and curvature-like components is properly determined by symmetry under index exchange, rather than by whether the tensor components are real or imaginary.

\begin{figure}[tbp]
\centering
\includegraphics[width=\columnwidth]{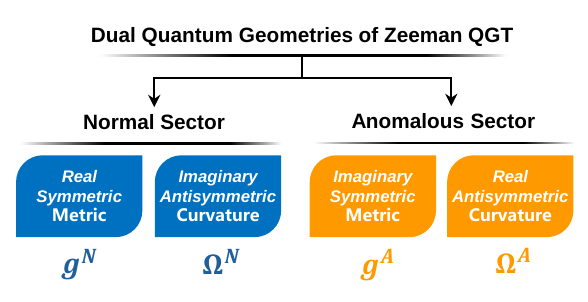}
\caption{Schematic illustration of the fourfold decomposition of Zeeman quantum geometric tensor. The metric components ${\bm g}^{N}$ and ${\bm g}^{A}$, and curvature components ${\bm \Omega}^{N}$ and ${\bm \Omega}^{A}$, are determined by symmetry or antisymmetry under index exchange, not by real or imaginary. }\label{Fig1}
\end{figure}

The conventional QGT $T^{ab}_{nm}$ is Hermitian, $T^{ab}_{nm} = (T^{ba}_{nm})^*$, which ensures that its imaginary part, the Berry curvature, is antisymmetric in indices $a$ and $b$, forming a pseudovector. Its real part, the quantum metric, is symmetric. Consequently, the linear conductivities arising from BC and QM inherit the same symmetry properties. In contrast, the Zeeman QGT does not satisfy this conjugate symmetry. As a result, the imaginary part is generally not antisymmetric in $a$ and $b$, and therefore does not exhibit the typical characteristic of a conventional Berry curvature.
By decomposing the Zeeman QGT into its symmetric and antisymmetric components, one obtains four distinct geometric quantities: two symmetric tensors (generalized Zeeman quantum metrics) and two antisymmetric tensors (Berry curvature-like quantities). In general, we define generalized Zeeman quantum 
geometric quantities as
\begin{eqnarray}
{\bm g}^{N/A,ab}_{nm} &=& \frac{1}{2}{\rm Re}/{\rm Im}[{ r}^a_{nm} { \sigma}^b_{mn}+{ \sigma}^a_{mn}{ r}^b_{nm} ], \nonumber\\
{\bf \Omega}^{A/N}_{nm} &=& - {\rm Re}/{\rm Im}({\bf r}_{nm} \times {\boldsymbol \sigma}_{mn}).\label{definition}
\end{eqnarray}
Hence ${\bm g}^{N/A}$ are metric-like (symmetric tensors) and ${\bf \Omega}^{A/N}$ are Berry curvature-like (polar vector), where $\Omega_c =(1/2) \epsilon_{cab} \Omega^{ab}$ is the vector representation of the antisymmetric tensor.

Replacing $\boldsymbol \sigma_{nm}\!\to\!\mathbf r_{nm}$ in Eq.~\eqref{definition} recovers the conventional case:
$\bm g^{N}_{nm}$ and $\boldsymbol\Omega^{N}_{nm}$ reduce to the standard quantum metric and Berry curvature, while
$\bm g^{A}_{nm}=\boldsymbol\Omega^{A}_{nm}=0$. We therefore refer to $(\bm g^{N},\boldsymbol\Omega^{N})$ as the
\textit{normal} Zeeman quantum metric/Berry curvature and to $(\bm g^{A},\boldsymbol\Omega^{A})$ as the \textit{anomalous} pair.
This fourfold decomposition of Zeeman QGT is illustrated in Fig.~\ref{Fig1}, which leads to a richer set of quantum responses and physical phenomena. Symmetry analysis further shows that the four quantities transform differently under crystalline and magnetic symmetries, allowing one to distinguish their individual roles in transport and optical responses.

\section{Hodge duals in 2D}
As pointed out in Ref.~[\onlinecite{XiangZeeman}], the normal ZBC satisfies
$\boldsymbol{\Omega}^{N}_n=\nabla_{\mathbf k}\times \boldsymbol{\sigma}_n/2$,
in direct analogy to the conventional Berry curvature ${\bf \Omega}_n = \nabla_k \times {\bf A}_n$ and thus plays a similar role in charge Hall responses driven by time-dependent magnetic fields. Below we show that the anomalous ZBC, $\boldsymbol{\Omega}^{A}$, likewise admits a Berry-curvature-like interpretation in two dimensions: it defines a local topological invariant and is the Hodge dual\cite{Azizi2025} of the winding field. For an in-plane vector ${\bf v}=(v_x,v_y)$, the two-dimensional (2D) Hodge star is $*{\bf v} \equiv \hat{\mathbf z}\times (v_x,v_y)=(-v_y,v_x)$~\cite{Gustafsson2022}, i.e., a $90^\circ$ rotation.~\cite{Gustafsson2022}

To make the discussion concrete, we consider a Dirac model $H = {\bf d}({\bf k}) \cdot {\boldsymbol \sigma}$ and examine the contribution of each component of Zeeman quantum geometry at low temperature. Define $\vect{v}_a \equiv \partial_a {\bf d}$ and $\vect{v}_{a,\perp}\equiv \vect{v}_a-\hat{\vect{d}}\bigl(\hat{\vect{d}} \cdot \vect{v}_a\bigr)$, we have the following general formulas for Zeeman quantum metrics (see \hyperref[app:zeeman-qgt]{Appendix A})
\begin{align}
{\bf \Omega}^N_n&=\;\frac{n}{2}\,\nabla_{\!k}\times \hat{\bm d},
\label{eq:N1}\\
{ g}^{A,ab}_n&=
-\frac{n}{4}\left[\partial_{k_a}\hat d_b+\partial_{k_b}\hat d_a\right],
\label{eq:N2}\\
{\bf \Omega}^A_n&=-\frac12\Big[\bd\,(\vk\!\cdot\!\bd)-(\bd\!\cdot\!\vk)\,\bd\Big],
\label{eq:N3}\\
{ g}^{N,ab}_n&= \frac14\left[(\bd\times \partial_{k_a}\bd)_b+(\bd\times \partial_{k_b}\bd)_a\right],
\label{eq:N4}
\end{align}
where $n=\pm$ labels the band. Note that ${\bf \Omega}^N_n$ and ${g}^{A,ab}_n$ depend on band index while ${\bf \Omega}^A_n$ and ${ g}^{N,ab}_n$ do not. As a consequence $\sum_n {\bf \Omega}^N_n=0$ but $\sum_n {\bf \Omega}^A_n \ne 0$. Using ${\boldsymbol\sigma}_n=n\,\hat{\mathbf d}$, Eq.~\eqref{eq:N1} reproduces $\boldsymbol{\Omega}^{N}_n=\nabla_{\mathbf k}\!\times\!\boldsymbol{\sigma}_n/2$. Hence the normal ZBC is divergence-free except at $\mathbf k=0$, inheriting the usual Berry-curvature singularity.
By contrast, the anomalous ZBC is generically neither divergence-free nor curl-free in 3D.

Focusing on anomalous ZBC $\boldsymbol{\Omega}^{A}$, we consider a massive 2D Dirac model
\begin{equation}
H=\mathbf d\cdot\boldsymbol{\sigma},\qquad \mathbf d=(k_x,k_y,m),
\label{ham}
\end{equation}
which becomes gapless in the massless limit \(m\to 0\).
From Eq.~(\ref{eq:N3}) one finds
$\boldsymbol{\Omega}^{A}=-\tfrac12 (k_x,k_y,2m)/d^2$ and
$\nabla_{\mathbf k}\!\cdot\!\boldsymbol{\Omega}^{A}=-m^2/d^4$ with $d^2=k_x^2+k_y^2+m^2$.
In the massless limit,
\begin{equation}
\nabla_{\mathbf k}\!\cdot\!\boldsymbol{\Omega}^{A}
=-\frac{1}{2}\,\nabla_{\mathbf k}\!\cdot\!\Big(\frac{\mathbf k}{k^2}\Big)
=C\,\delta^{(2)}(\mathbf k),\label{eq9}
\end{equation}
where $C=-\pi$ is the flux carried by the 2D ``monopole'', which represents the 2D point-source charge for ${\bf \Omega}^A$. Eq.~(\ref{eq9}) shows that \(\boldsymbol{\Omega}^{A}\) defines a geometric flux field associated with the Dirac node.

From Eq.~(\ref{eq9}), one can define a Gauss-type topological index as
\begin{equation}
Q=\frac{1}{C}\iint_S \nabla_{\mathbf k}\!\cdot\!\boldsymbol{\Omega}^{A}\,d^2k
=\frac{1}{C}\oint_{\partial S}\boldsymbol{\Omega}^{A}\!\cdot\!\hat{\mathbf n}\,dl=1,
\end{equation}
where ${\hat n}$ is the outward unit normal to the boundary curve $\partial S$ in the plane. Thus Eq.~(\ref{eq9}) characterizes the node by a quantized local topological charge. This provides a flux-based description of the Dirac singularity.

\begin{figure}[tbp]
\centering
\includegraphics[width=\columnwidth]{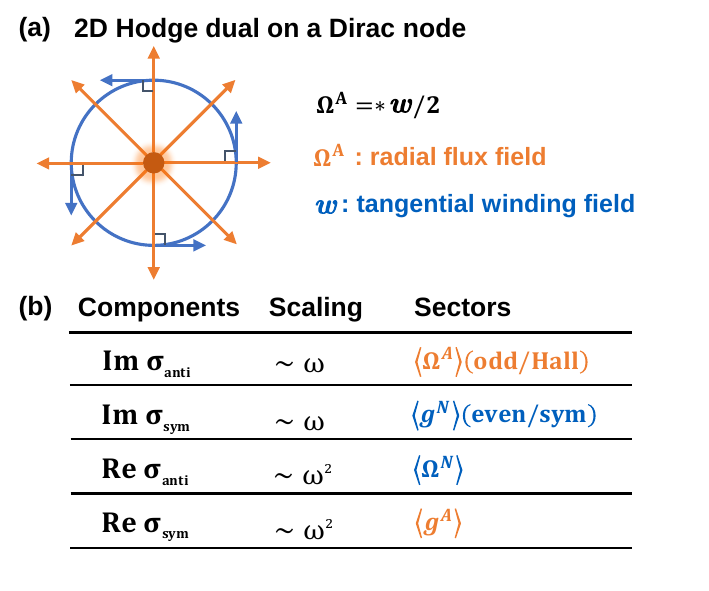}
\caption{ (a) Schematic plot of the 2D Hodge dual between anomalous Zeeman Berry curvature ${\boldsymbol \Omega}^{A}$ and the winding field ${\boldsymbol w}$. (b) One-to-one correspondence between the sectors of Zeeman QGT and frequency-resolved gyrotropic conductivity.}\label{Fig2}
\end{figure}

The same node also admits the familiar winding description.
The winding number is defined as\cite{Montambaux2018}
\begin{equation}
C_w=\frac{1}{2\pi}\oint_\Gamma \nabla_{\mathbf k}\theta_{\mathbf k}\cdot d\mathbf l,
\qquad \tan\theta_{\mathbf k}=\frac{d_y}{d_x}.
\end{equation}
From Eq.~(\ref{ham}), the corresponding winding field is ${\boldsymbol w} \equiv \nabla_{\mathbf k}\theta_{\mathbf k}=(-k_y,k_x)/k^2$, which is everywhere tangential and divergence-free. By contrast, $-2\,\boldsymbol{\Omega}^{A}$ is radial and curl-free in the massless limit. Although their local geometries are different, both fields encode the same singularity and yield the same topological charge.
Hence, in 2D they are Hodge duals: one, \(\boldsymbol{\Omega}^{A}\), provides a flux (Gauss-type) index, while the other, \(\boldsymbol{w}\), provides a vortex (winding) index, as shown in Fig.~\ref{Fig2}(a).

This leads to the general 2D Hodge duality statement: for any two-band Hamiltonian $H = {\bf d} \cdot {\boldsymbol \sigma}$ whose ${\bf d}$-vector is planar on a 2D $k$-domain (i.e. $d_z = 0$), the anomalous Zeeman flux field is Hodge-dual to the winding field (see \hyperref[app:zeeman-qgt]{Appendix A}) with the relation
${\bf \Omega}^A = *{\boldsymbol w}/2$ where the Hodge star symbol $"*"$ rotates planar vectors by $\pi/2$.
Correspondingly,
\begin{equation}
[\nabla_{\!k}\!\times\!\boldsymbol{\Omega}^{A}]_z=\frac{1}{2}\,\nabla_{\!k} \cdot {\boldsymbol w} ,~~~
\nabla_{\!k}\!\cdot\!\boldsymbol{\Omega}^{A}=-\frac{1}{2}\,[\nabla_{\!k}\!\times\!{\boldsymbol w}]_z. \label{HodgeDual}
\end{equation}

For the canonical vortex field \({\boldsymbol w}=\nabla_{\mathbf k}\theta_{\mathbf k}\sim \hat{\boldsymbol\theta}/k\), Eq.~(\ref{HodgeDual}) gives
\[
[\nabla_{\mathbf k}\times {\boldsymbol w}]_z = 2\pi \delta^{(2)}(\mathbf k),\qquad
\nabla_{\mathbf k}\cdot {\boldsymbol w} = 0.
\]
Its Hodge-dual field \(\boldsymbol{\Omega}^{A}=\ast {\boldsymbol w}/2\) is instead radial and satisfies
\[
\nabla_{\mathbf k}\cdot \boldsymbol{\Omega}^{A} = -\pi \delta^{(2)}(\mathbf k),\qquad
[\nabla_{\mathbf k}\times \boldsymbol{\Omega}^{A}]_z = 0.
\]
Thus, the same isolated singularity may be described either as a tangential vortex or, equivalently, as a radial flux source.

More generally, for any isolated 2D winding defect with
\begin{equation}
\oint_{\Gamma} {\boldsymbol w}\cdot d\mathbf l = 2\pi C_w,\label{win1}
\end{equation}
where $C_w$ denotes the winding number and \(\Gamma\) is any positively oriented closed contour enclosing the defect, one may define the anomalous Zeeman BC as
\begin{equation}
\boldsymbol{\Omega}^A=\frac{1}{2}\,\hat{\mathbf z}\times {\boldsymbol w} .\label{dual2}
\end{equation}
It then follows from Eqs.~(\ref{win1})-(\ref{dual2}) that
\[
\oint_{\Gamma}\boldsymbol{\Omega}^{A}\cdot \hat{\mathbf n}\,dl = -\pi C_w,
\qquad
\nabla_{\mathbf k}\cdot \boldsymbol{\Omega}^{A} = -\pi C_w\,\delta^{(2)}(\mathbf k).
\]
After normalization by \(-\pi\), the anomalous Zeeman BC therefore defines an integer 2D monopole charge equal to the winding number $C_w$. This shows that the vortex-flux correspondence is not specific to the Dirac model, but is the universal local structure of an isolated 2D winding singularity.

Besides the Gauss-type index and the winding number, the Dirac node can also be characterized by another local topological invariant, namely the Berry phase. For this purpose, we introduce the Berry curvature
\begin{equation}
\Omega_{xy}=-\frac{n}{2}(\partial_{k_x}\hat{\mathbf d}\times\partial_{k_y}\hat{\mathbf d})\cdot\hat{\mathbf d} \label{eq10}
\end{equation}
with $\hat{\mathbf d} = {\bf d}/d$. For Eq.~\eqref{ham}, one finds
$\Omega_{xy}=-nm\big/ \big[2\,(k_x^2+k_y^2+m^2)^{3/2}\big]$. In the $m\!\to\!0$ limit, we obtain $\Omega_{xy} = -n \pi {\rm sgn}(m) \delta^{(2)}({\bf k})$.
While $\Omega_{xy}\neq \nabla_{\mathbf k}\cdot\boldsymbol{\Omega}^{A}$ pointwise, their integrals over any disk enclosing the node agree up to a sign in the $m\!\to\!0$ limit, showing that they capture the same local topological charge.

\smallskip
Finally, it is useful to distinguish these local nodal invariants from the global band topology encoded by the conventional Berry curvature. The winding field ${\boldsymbol w}$ characterizes the local topology of a Dirac node, whereas the nontrivial global topology of its gapped counterpart is labeled by the Chern number, which governs the quantized Hall effect. Within Zeeman quantum geometry, the anomalous geometric flux field ${\boldsymbol \Omega}^{A}$ arises naturally and provides a complementary description of the same Dirac singularity in terms of measurable spin or pseudospin textures and interband mode couplings.
Thus, ${\boldsymbol w}$ and ${\boldsymbol \Omega}^{A}$ provide Hodge-dual descriptions of the same local Dirac singularity: the former as a tangential vortex and the latter as a radial flux source. In this sense, local $\pi_1$ topology can be represented by a curvature-flux invariant, in parallel with the curvature representation of global $\pi_2$ topology by the conventional Berry curvature.
As we show below, this normal-anomalous decomposition of Zeeman QGT is directly reflected in gyrotropic transport.

\section{Manifestations in 2D Transport}
In the absence of additional symmetry constraints, all four components of the Zeeman quantum geometric tensor can, in principle, contribute to the gyrotropic response. A time-dependent magnetic field induces a linear charge current,
\begin{eqnarray}
J_a = \sigma_{ab} B_b\nonumber,
\end{eqnarray}
which defines the frequency-dependent gyrotropic conductivity tensor. Because a static magnetic field does not induce a current, we remove the dc contribution and obtain (see \hyperref[app:linear-response]{Appendix B})
\begin{eqnarray}
{\bf J}(\omega)
&=& g\mu_B \sum_{nm} \int_k f_n[\frac{i\omega }{\epsilon_{nm}} {\bf B} \times {\bf \Omega}^A_{nm} -i\frac{2\omega }{\epsilon_{nm}} {\bm g}^N_{nm} \cdot {\bf B}\nonumber\\
&-& \frac{\omega^2 }{\epsilon_{nm}^2} {\bf B} \times {\bf \Omega}^N_{nm} + 2\frac{\omega^2 }{\epsilon_{nm}^2} {\bm g}^A_{nm} \cdot {\bf B}], \label{exp}
\end{eqnarray}
where $g$ is the $g$-factor, $\mu_B$ is the Bohr magneton, $f_n$ is the equilibrium Fermi distribution function, and $\epsilon_{nm}=\epsilon_n-\epsilon_m$ is the energy difference.
To resolve the tensor structure, we decompose the transverse conductivity into antisymmetric and symmetric parts,
\begin{eqnarray}
\sigma^{-} = \frac{1}{2} (\sigma_{xy} - \sigma_{yx}), ~~~ \sigma^{+} = \frac{1}{2} (\sigma_{xy} + \sigma_{yx}),\label{gyroconduc}
\end{eqnarray}
where $\sigma^-$ is the Hall component and $\sigma^+$ is the symmetric transverse component. One then obtains the low-frequency selection rules
\begin{eqnarray}
{\rm Im}~\sigma^{-}(\omega) &\sim& \omega \langle {\bf \Omega}^A \rangle, ~~~ {\rm Im}~\sigma^{+}(\omega) \sim \omega \langle {\bm g}^N \rangle, \nonumber\\
{\rm Re}~\sigma^{-}(\omega) &\sim& \omega^2 \langle {\bf \Omega}^N \rangle, ~~~ {\rm Re}~\sigma^{+}(\omega) \sim \omega^2 \langle {\bm g}^A \rangle, \label{rule}
\end{eqnarray}
where $\langle ... \rangle$ denotes the Brillouin-zone weighted integration. Eq.~(\ref{rule}) makes the one-to-one correspondence transparent: the odd-$\omega$ antisymmetric channel probes ${\bf \Omega}^A$, the odd-$\omega$ symmetric channel probes ${\bm g}^N$, the subleading $O(\omega^2)$ real Hall channel probes ${\bf \Omega}^N$, and the subleading $O(\omega^2)$ real symmetric channel probes ${\bm g}^A$. Therefore, the combination of tensor symmetry and low-frequency scaling provides a practical route to disentangle the normal and anomalous sectors of Zeeman quantum geometry.

Equation (\ref{rule}) gives an operational protocol for detecting the anomalous Zeeman curvature. In magnetic point groups where the symmetric transverse channels are forbidden, the leading imaginary Hall response $\operatorname{Im}\sigma^{-}(\omega)\propto \omega\langle\Omega^{A}\rangle$ provides a direct transport signature of the anomalous curvature sector. Conversely, groups that suppress $\Omega^{A}$ but allow $g^{N}$ isolate the normal metric sector. The symmetry restrictions on these response channels follow from Neumann's principle and are summarized in Table I. For example, the real Hall response associated with $\Omega^{N}$ is allowed in all magnetic point groups, while groups such as $3m$, $4mm$, and $6mm$ additionally permit an $O(\omega)$ Hall response governed by $\Omega^{A}$ and forbid the symmetric transverse responses associated with $g^{N}$ and $g^{A}$. By contrast, groups such as $m' m'2$, $3m'$, $4m'm'$, $4'm'm$, and $6m'm'$ allow only the symmetric transverse response associated with $g^{N}$.

\begin{table}[t!]
\centering
\begin{tabular}{ccc}
\hline\hline
& Jahn's notation & Symmetry allowed\\
\hline
$g^N$ & $ae[V^2]$ &
\makecell[c]{$\textcolor{blue}{1,m,2,2',mm2,m'm2',3,4,6,}$\\
 $m'm'2, 3m',4',4m'm',4'm'm,6m'm'$}\\
\hline
$g^A$ & $e[V^2]$  &\makecell[c]{$\textcolor{blue}{1,m,2,2',mm2,m'm2',3,4,6,}$\\$11',m',m1',21', mm21',m'm'2,$\\$31', 41',4',61',6'$}\\
\hline
$\Omega^N$ & $e\{V^2\}$ &All\\
\hline
$\Omega^A$ & $ae\{V^2\}$ &
\makecell[c]{$\textcolor{blue}{1,m,2,2',mm2,m'm2',3,4,6,}$\\$m',3m,4mm,6mm$}\\
\hline\hline
\end{tabular}
\caption{Magnetic point-group classification of the four Zeeman quantum geometric sectors in two-dimensional gyrotropic transport. For each 2D magnetic point group, the table indicates whether the corresponding conductivity channel associated with $\Omega^{N}$, $\Omega^{A}$, $g^{N}$, or  $g^{A}$ is symmetry allowed. The leading low-frequency responses follow Eq.~(\ref{rule}): $\Omega^{A}$ and $g^{N}$ appear at $O(\omega)$, whereas $\Omega^{N}$ and $g^{A}$ appear at $O(\omega^2)$. The magnetic point groups highlighted in blue allow all four components of the Zeeman QGT.}
\label{tab:sigmaD_shape}
\end{table}



\section{Verification via reciprocal KME response}
Direct experimental access to the gyrotropic Hall effect (GHE) remains challenging. A more practical route is provided by its reciprocal response, the kinetic magnetoelectric effect (KME), which has already been observed experimentally in Refs.~[\onlinecite{Furukawa2017,Xu2020,Ni2021}]. In the present context, GHE refers to a charge current induced by a time-dependent magnetic field, whereas KME describes an induced magnetization generated by an electric field. By Onsager reciprocity, these two effects probe the same Zeeman-geometric sectors in reciprocal measurement configurations\cite{Zhong2016}. The KME therefore offers an experimentally accessible route to test the geometric decomposition established above.

From linear-response theory, the induced magnetization $\bf M$ under a time-dependent electric field takes the form (see \hyperref[app:linear-response]{Appendix B})
\begin{eqnarray}
{\bf M}
&=& \sum_{nm} \int_k \frac{f_n}{2\epsilon_{nm}}[\frac{i\omega }{\epsilon_{nm}} {\bf E} \times {\bf \Omega}^N_{mn} +\frac{2i\omega}{\epsilon_{nm}} {\bm g}^A_{mn} \cdot {\bf E}\nonumber\\
&+& \frac{\omega^2 }{\epsilon_{nm}^2} {\bf E} \times {\bf \Omega}^A_{mn} + \frac{2\omega^2 }{\epsilon_{nm}^2} {\bm g}^N_{mn} \cdot {\bf E}]. \label{exp1}
\end{eqnarray}
Here the frequency-independent contribution has been omitted, since the Onsager relation between the GHE and the KME holds only for the frequency-dependent contribution\cite{Zhong2016}. Eq.~(\ref{exp1}) shows that the reciprocal KME probes the same four Zeeman-geometric sectors as the gyrotropic conductivity, but in a complementary arrangement. Writing \(M_a=\alpha_{ab}E_b\), this complementary correspondence becomes transparent once the magnetoelectric tensor $\alpha_{ab}$ is resolved into four components.

Similar to Eq.~(\ref{gyroconduc}), we define
\begin{eqnarray}
\alpha^- = \frac{1}{2} (\alpha_{xy} - \alpha_{yx}), ~~~ \alpha^+ = \frac{1}{2} (\alpha_{xy} + \alpha_{yx}) \nonumber
\end{eqnarray}
as the antisymmetric Hall and symmetric transverse components, respectively. One then obtains the low-frequency selection rules
\begin{eqnarray}
{\rm Im}~\alpha^-(\omega) &\sim& \omega \langle {\bf \Omega}^N \rangle, ~~~ {\rm Im}~\alpha^+(\omega) \sim \omega \langle {\bm g}^A \rangle, \nonumber\\
{\rm Re}~\alpha^-(\omega) &\sim& \omega^2 \langle {\bf \Omega}^A \rangle, ~~~ {\rm Re}~\alpha^+(\omega) \sim \omega^2 \langle {\bm g}^N \rangle. \nonumber
\end{eqnarray}
Thus, in contrast to the gyrotropic conductivity, the reciprocal KME response probes the same four geometric sectors in a complementary arrangement: the leading $O(\omega)$ channels isolate ${\bf \Omega}^N$ and ${\bm g}^A$, while the subleading $O(\omega^2)$ channels isolate ${\bf \Omega}^A$ and
${\bm g}^N$. Therefore, combining GHE and KME measurements provides a symmetry- and frequency-resolved experimental strategy for disentangling the normal and anomalous sectors of Zeeman quantum geometry.

\section{Decomposition of the spin quantum geometric tensor}

The same normal--anomalous decomposition can be extended directly to the spin quantum geometric tensor introduced in Ref.~[\onlinecite{Xiang-spin}], where the antisymmetric sector yields both a conventional spin Berry curvature\cite{universal} and an anomalous real spin-vorticity contribution\cite{Mertig,W-Yao}. For a band pair $(n,m)$, we define the spin QGT as
\begin{eqnarray}
 (T^s)^{ab,c}_{nm} = v^{a,c}_{nm} v^b_{mn}/\epsilon_{nm}^2\label{sQGT}
\end{eqnarray}
where $c$ labels the spin-polarization direction, $v^b=\partial_{k_b}H$ is the velocity operator, and
\begin{equation}
v^{a,c}=\frac{1}{4}\left\{ v^a,\sigma^c \right\}
\end{equation}
is the spin-velocity operator. Replacing $v_{nm}^{a,c}$ by the ordinary velocity matrix element reduces Eq.~(\ref{sQGT}) to the conventional two-band QGT.

Following the tensor decomposition in Eq.~(\ref{definition}) of the main text, we separate the spin QGT into symmetric metric-like and antisymmetric curvature-like parts. The normal and anomalous spin quantum metrics are
\begin{equation}
(g_s^{N})_{nm}^{ab,c}
=
\frac{1}{2}\,
\mathrm{Re}
\left[
\frac{
v_{nm}^{a,c} v_{mn}^{b}
+
v_{nm}^{b,c} v_{mn}^{a}
}{\epsilon_{nm}^2}
\right],
\end{equation}
\begin{equation}
(g_s^{A})_{nm}^{ab,c}
=
\frac{1}{2}\,
\mathrm{Im}
\left[
\frac{
v_{nm}^{a,c} v_{mn}^{b}
+
v_{nm}^{b,c} v_{mn}^{a}
}{\epsilon_{nm}^2}
\right],
\end{equation}
while the normal and anomalous spin Berry-curvature-like quantities are
\begin{equation}
\bm{\Omega}_{s,nm}^{N,c}
=
-
\mathrm{Im}
\left[
\frac{
\bm{v}_{nm}^{\,c}\times \bm{v}_{mn}
}{\epsilon_{nm}^2}
\right],
\end{equation}
\begin{equation}
\bm{\Omega}_{s,nm}^{A,c}
=
-
\mathrm{Re}
\left[
\frac{
\bm{v}_{nm}^{\,c}\times \bm{v}_{mn}
}{\epsilon_{nm}^2}
\right].
\end{equation}
Here $\bm{v}_{nm}^{\,c}$ denotes the vector of spin-velocity matrix elements with spin polarization $c$. Thus, exactly as in the Zeeman case, spin quantum geometry contains four distinct sectors: the normal metric $g_s^N$, the anomalous metric $g_s^A$, the normal spin curvature $\bm{\Omega}_s^N$, and the anomalous spin curvature $\bm{\Omega}_s^A$.

For later use, we introduce the band-resolved geometric quantities
\begin{equation}
\bar{\bm{\Omega}}_n^{A,c}
=
\sum_m \epsilon_{nm}\,\bm{\Omega}_{s,nm}^{A,c},
\qquad
\bar{g}_n^{N,c}
=
\sum_m \epsilon_{nm}\,g_{s,nm}^{N,c}.\label{bar}
\end{equation}
In particular, the anomalous spin-curvature sector satisfies
\begin{equation}
\bar{\bm{\Omega}}_n^{A,c}
=
\frac{1}{2}\,\nabla_{\bm{k}}\times \bm{v}_n^{\,c},
\end{equation}
which has the same formal structure as an ordinary Berry curvature, with the spin velocity $\bm{v}_n^{\,c}$ playing the role of an effective connection. This quantity is precisely the geometric object closely related to the spin vorticity discussed in Refs.~[\onlinecite{Mertig}]-[\onlinecite{W-Yao}].

The electric-field-induced spin current derived in \hyperref[app:spin-current]{Appendix C} can be organized in terms of the four sectors of spin QGT. Writing $\int_{\bm{k}} \equiv \int_{\mathrm{BZ}} d^2k/(2\pi)^2$, the band-resolved contribution takes the form
\begin{equation}
\bm{J}_n^{\,c}
=
\int_{\bm{k}} f_n
\left[
\bm{E}\times
\left({\bm{\Omega}}_n^{N,c}
+
\tau \bar{\bm{\Omega}}_n^{A,c}
\right)
-
2\left({g}_n^{A,c}
+
\tau \bar{g}_n^{N,c}
\right)\cdot \bm{E}
\right],
\label{eq:spin_current_appendixA}
\end{equation}
where $\tau$ is the relaxation time. Eq.~(\ref{eq:spin_current_appendixA}) makes the physical content of the decomposition transparent: the antisymmetric response channels are governed by the spin-curvature sectors, whereas the symmetric channels are governed by the spin-metric sectors.

Equation~(\ref{eq:spin_current_appendixA}) reveals a clear one-to-one correspondence between the four sectors of the spin QGT and their distinct contributions to the spin current. In particular, the normal and anomalous spin-curvature terms, \({\bm{\Omega}}_n^{N,c}\) and \(\bar{\bm{\Omega}}_n^{A,c}\), govern the Hall-like response, with the two contributions exhibiting different $\tau$ scaling. Likewise, the anomalous and normal spin-metric terms, \({g}_n^{A,c}\) and \(\bar{g}_n^{N,c}\), determine the symmetric response, again with distinct $\tau$-dependence. This decomposition makes it transparent that each spin-transport channel can be traced directly and unambiguously to its corresponding spin geometric origin.

In short, spin QGT also exhibits a normal--anomalous decomposition parallel to that of the Zeeman QGT. The normal sector contains the conventional spin Berry-curvature contribution, whereas the anomalous sector gives rise to a real spin-vorticity-like field\cite{Mertig}. In systems with chiral symmetry, the transverse response from the metric sector can be suppressed, leaving a purely Hall-type spin response\cite{W-Yao}. These results show that the geometric structure extends naturally beyond Zeeman coupling and provides a unified framework for charge, spin, and vorticity-like transport responses.

\section{Summary}
We have shown that the non-Hermitian Zeeman QGT contains a normal--anomalous metric-curvature structure determined by symmetry under index exchange. The normal sector reduces to the conventional Hermitian QGT, while the anomalous sector contains an imaginary symmetric metric-like tensor and a real antisymmetric curvature-like tensor absent in the conventional QGT. In a two-dimensional Dirac system, the anomalous Zeeman curvature forms a radial flux field that is Hodge-dual to the tangential winding field of the Dirac node. It therefore converts local $\pi_1$ winding topology into a Gauss-type curvature-flux invariant. We further show that the four Zeeman-geometric sectors are mapped onto distinct gyrotropic conductivity channels resolved by tensor symmetry and low-frequency scaling, and that the reciprocal kinetic magnetoelectric response provides a complementary probe. These results establish non-Hermitian Zeeman quantum geometry as a measurable framework connecting local Dirac-node topology with transport signatures.

\section*{Acknowledgement}
This work was supported by the National Natural Science Foundation of China (Grants No.~12404059 and No.~12574054).

\newpage
\onecolumngrid

\phantomsection
\label{app:zeeman-qgt}
\section*{Appendix A: Zeeman quantum metrics for the Dirac model}
Consider a two-band Pauli Hamiltonian~\cite{Liu2024}
\begin{equation}
    H(\vect{k}) = \vect{d}(\vect{k}) \cdot \bm{\sigma}, \qquad
    \vect{d} = (d_x, d_y, d_z).
\end{equation}
with eigenvalues \(\varepsilon_\pm = \pm \mathcal{E}\), where
\(\mathcal{E}\equiv|\vect{d}|\), and
\(\hat{\vect{d}}=\vect{d}/\mathcal{E}\). The projectors are
\begin{equation}
    P_\pm = \frac{1 \pm \hat{\vect{d}} \cdot \bm{\sigma}}{2}.
\end{equation}
We denote the derivative of \(H\) with respect to \(k_a\) by
\begin{equation}
   \partial_{k_a} H = (\partial_{k_a} \vect{d}) \cdot \bm{\sigma}
    \equiv \vect{v}_a \cdot \bm{\sigma}
    \quad (a \in \{x, y\}).
\end{equation}
The interband Berry connection matrix element is~\cite{Xiao2010}
\begin{equation}
    r^a_{nm} = i \frac{\langle u_n | \partial_{k_a} H | u_m\rangle}{\varepsilon_m - \varepsilon_n}
    = \frac{i}{\varepsilon_m - \varepsilon_n} \langle u_n| \vect{v}_a \cdot \bm{\sigma} | u_m \rangle \quad (m \neq n).
\end{equation}
We can define the band-diagonal object
\begin{equation}
    {T}^{Z,ab}_{n} = \sum_{m \neq n} r^a_{nm} \, \sigma^b_{mn}, \qquad
    \sigma^b_{mn} = \langle u_m | \sigma^b | u_n \rangle.
\end{equation}

\subsection{Universal two-band identity for \({T}^{Z,ab}_{n}\)}

In this section, we derive the projector-trace form of \({T}^{Z,ab}_{n}\). We start from a two-band model with eigenpairs \(H|u_{\pm}\rangle = \varepsilon_{\pm}|u_{\pm}\rangle\) and projectors \(P_{\pm} = |u_{\pm}\rangle\langle u_{\pm}|\). The starting point is
\begin{equation}
    {T}^{Z,ab}_{n} = \sum_{m \neq n} r^{a}_{nm} \sigma^b_{mn}, \qquad
    r^{a}_{nm} = i \frac{\langle u_n| \partial_{k_a} H |u_m\rangle}{\varepsilon_m - \varepsilon_n}, \qquad
    \sigma^b_{mn} = \langle u_m|\sigma^b|u_n\rangle. \label{supp:eq:B1}
\end{equation}
For a two-band model, the sum reduces to the opposite band \(m = \bar n\). Denoting \(\Delta\varepsilon \equiv \varepsilon_{\bar n} - \varepsilon_n\), we have
\begin{align}
    {T}^{Z,ab}_{n}
    &= i \frac{\langle u_n| (\partial_{k_a} H) |u_{\bar n}\rangle \langle u_{\bar n}|\sigma^b|u_n\rangle}{\Delta\varepsilon}
    = i \frac{\langle u_n| (\partial_{k_a} H) P_{\bar n} \sigma^b |u_n\rangle}{\Delta\varepsilon}\nonumber \\
    &= i \frac{\Tr\!\left[ |u_n\rangle\langle u_n| (\partial_{k_a} H) P_{\bar n} \sigma^b \right]}{\Delta\varepsilon}
    = i \frac{\Tr\!\left[ P_n (\partial_{k_a} H) P_{\bar n} \sigma^b \right]}{\Delta\varepsilon}. \label{supp:eq:B2}
\end{align}
Since \(\partial_{k_a} H = v_a \cdot \bm{\sigma}\), we obtain the projector-trace representation
\begin{equation}
{T}^{Z,ab}_{n} = i \frac{\Tr\!\left[ P_n (v_a \cdot \bm{\sigma}) P_{\bar n} \sigma^b \right]}{\varepsilon_{\bar n} - \varepsilon_n}. \label{supp:eq:B3}
\end{equation}

For a multiband generalization, the same structure is obtained by replacing \(P_{\bar n}/(\varepsilon_{\bar n} - \varepsilon_n)\) with the reduced resolvent \(G'_n = \sum_{m \neq n} P_m/(\varepsilon_m - \varepsilon_n)\).

Recall that
\begin{equation}
    P_n = \frac{1 + n \hat{\vect{d}} \cdot \bm{\sigma}}{2}, \qquad
    P_{\bar n} = \frac{1 - n \hat{\vect{d}} \cdot \bm{\sigma}}{2}, \qquad
    \vect{v}_a = \partial_{k_a} \vect{d}, \quad \vect{v}_a \cdot \bm{\sigma} = \sum_i (v_a)_i \sigma_i,
\end{equation}
we have
\begin{equation}
    P_n (\vect{v}_a \cdot \bm{\sigma}) P_{\bar n}
    = \frac{1}{4} (1 + n \hat{\vect{d}} \cdot \bm{\sigma}) (\vect{v}_a \cdot \bm{\sigma}) (1 - n \hat{\vect{d}} \cdot \bm{\sigma}).
\end{equation}

In the following, we first evaluate the two-factor and three-factor products appearing in this expression, and then assemble these results to obtain the final result.

\noindent{1. Two-factor product.}

Using \(\sigma_i \sigma_j = \delta_{ij} \mathbb{I} + i \eps_{ijk} \sigma_k\), for any real vectors \(\vect{A}, \vect{B}\),
\begin{equation}
    (\vect{A} \cdot \bm{\sigma}) (\vect{B} \cdot \bm{\sigma})
    = (\vect{A} \cdot \vect{B}) \mathbb{I} + i (\vect{A} \times \vect{B}) \cdot \bm{\sigma}.
\end{equation}
Therefore,
\begin{align}
    (\hat{\vect{d}} \cdot \bm{\sigma}) (\vect{v}_a \cdot \bm{\sigma})
    &= \hat{\vect{d}} \cdot \vect{v}_a \, \mathbb{I} + i (\hat{\vect{d}} \times \vect{v}_a) \cdot \bm{\sigma}, \label{supp:eq:A1} \\
    (\vect{v}_a \cdot \bm{\sigma}) (\hat{\vect{d}} \cdot \bm{\sigma})
    &= \hat{\vect{d}} \cdot \vect{v}_a \, \mathbb{I} - i (\hat{\vect{d}} \times \vect{v}_a) \cdot \bm{\sigma}. \label{supp:eq:A2}
\end{align}

\noindent{2. Three-factor product.}

The three-factor term is
\begin{align}
(\hat{\vect{d}} \cdot \bm{\sigma}) (\vect{v}_a \cdot \bm{\sigma}) (\hat{\vect{d}} \cdot \bm{\sigma})
    &= \big( \hat{\vect{d}} \cdot \vect{v}_a + i (\hat{\vect{d}} \times \vect{v}_a) \cdot \bm{\sigma} \big) (\hat{\vect{d}} \cdot \bm{\sigma})\nonumber \\
    &= (\hat{\vect{d}} \cdot \vect{v}_a) (\hat{\vect{d}} \cdot \bm{\sigma}) + i \big[ (\hat{\vect{d}} \times \vect{v}_a) \cdot \bm{\sigma} \big] (\hat{\vect{d}} \cdot \bm{\sigma}). \label{supp:eq:A3}
\end{align}
Since \((\hat{\vect{d}} \times \vect{v}_a) \cdot \hat{\vect{d}} = 0\), and
\begin{equation}
    (\hat{\vect{d}} \times \vect{v}_a) \times \hat{\vect{d}} = \vect{v}_a - \hat{\vect{d}} (\hat{\vect{d}} \cdot \vect{v}_a) \equiv \vect{v}_{a,\perp},
\end{equation}
we obtain
\begin{equation}
    \big[ (\hat{\vect{d}} \times \vect{v}_a) \cdot \bm{\sigma} \big] (\hat{\vect{d}} \cdot \bm{\sigma})
    = i \, \vect{v}_{a,\perp} \cdot \bm{\sigma}. \label{supp:eq:A4}
\end{equation}
Therefore,
\begin{equation}
    (\hat{\vect{d}} \cdot \bm{\sigma}) (\vect{v}_a \cdot \bm{\sigma}) (\hat{\vect{d}} \cdot \bm{\sigma})
    = (\hat{\vect{d}} \cdot \vect{v}_a) (\hat{\vect{d}} \cdot \bm{\sigma}) - \vect{v}_{a,\perp} \cdot \bm{\sigma}. \label{supp:eq:A5}
\end{equation}

\noindent{3. Assemble the pieces.}

Expanding
\begin{align}
    P_n (\vect{v}_a \cdot \bm{\sigma}) P_{\bar n}
    &= \frac{1}{4} \Big[ (\vect{v}_a \cdot \bm{\sigma})
    - n (\vect{v}_a \cdot \bm{\sigma}) (\hat{\vect{d}} \cdot \bm{\sigma})
    + n (\hat{\vect{d}} \cdot \bm{\sigma}) (\vect{v}_a \cdot \bm{\sigma})
    - (\hat{\vect{d}} \cdot \bm{\sigma}) (\vect{v}_a \cdot \bm{\sigma}) (\hat{\vect{d}} \cdot \bm{\sigma}) \Big] \nonumber\\
    &= \frac{1}{4} \Big[ (\vect{v}_a \cdot \bm{\sigma})
    + 2 i n (\hat{\vect{d}} \times \vect{v}_a) \cdot \bm{\sigma}
    - (\hat{\vect{d}} \cdot \vect{v}_a) (\hat{\vect{d}} \cdot \bm{\sigma})
    + \vect{v}_{a,\perp} \cdot \bm{\sigma} \Big]. \label{supp:eq:A6}
\end{align}
Since \((\vect{v}_a \cdot \bm{\sigma}) - (\hat{\vect{d}} \cdot \vect{v}_a) (\hat{\vect{d}} \cdot \bm{\sigma}) = \vect{v}_{a,\perp} \cdot \bm{\sigma}\), we conclude
\begin{equation}
    P_n (\vect{v}_a \cdot \bm{\sigma}) P_{\bar n}
    = \frac{1}{2} \Big[ \vect{v}_{a,\perp} + i n (\hat{\vect{d}} \times \vect{v}_a) \Big] \cdot \bm{\sigma}. \label{supp:eq:A7}
\end{equation}

Substituting Eq.\eqref{supp:eq:A7} into the trace in Eq.\eqref{supp:eq:B3} gives
\begin{equation}
    \Tr[ P_n (\vect{v}_a \cdot \bm{\sigma}) P_{\bar n} \sigma^b ] = \Tr\!\left[ \frac{1}{2} \big[ \vect{v}_{a,\perp} + i n (\hat{\vect{d}} \times \vect{v}_a) \big] \cdot \bm{\sigma} \, \sigma^b \right]. \label{supp:eq:B5}
\end{equation}
Using the Pauli matrix trace properties \(\Tr[\sigma^i] = 0\), \(\Tr[\sigma^i \sigma^j] = 2 \delta_{ij}\), and for any vector \(\vect{A}\): \(\Tr[(\vect{A} \cdot \bm{\sigma}) \sigma^b] = 2 A_b\), we find
\begin{align}
    \Tr[(\vect{v}_{a,\perp} \cdot \bm{\sigma}) \sigma^b] &= 2 (\vect{v}_{a,\perp})_b, \label{supp:eq:B6} \\
    \Tr[i n (\hat{\vect{d}} \times \vect{v}_a) \cdot \bm{\sigma} \, \sigma^b] &= 2 i n (\hat{\vect{d}} \times \vect{v}_a)_b. \label{supp:eq:B7}
\end{align}
Thus,
\begin{equation}
    \Tr[ P_n (\vect{v}_a \cdot \bm{\sigma}) P_{\bar n} \sigma^b ] = \frac{1}{2} \big[ 2 (\vect{v}_{a,\perp})_b + 2 i n (\hat{\vect{d}} \times \vect{v}_a)_b \big] = (\vect{v}_{a,\perp})_b + i n (\hat{\vect{d}} \times \vect{v}_a)_b. \label{supp:eq:B8}
\end{equation}
For the two-band spectrum, the energy denominator is
\begin{equation}
    \varepsilon_{\bar n} - \varepsilon_n = -2n\mathcal{E}. \label{supp:eq:B9}
\end{equation}

Combining the above results, we finally obtain
\begin{equation}
    {T}^{Z,ab}_{n} = i \frac{(\vect{v}_{a,\perp})_b + i n (\hat{\vect{d}} \times \vect{v}_a)_b}{-2n\mathcal{E}} = -\frac{i}{2n\mathcal{E}} (\vect{v}_{a,\perp})_b + \frac{1}{2\mathcal{E}} (\hat{\vect{d}} \times \vect{v}_a)_b. \label{supp:eq:B10}
\end{equation}

We decompose \({T}^{Z,ab}_{n}\) into normal and anomalous Zeeman quantum geometric quantities according to Eq.~(3) of the main text
\begin{align}
    g^{N,ab}_{n} &\equiv \frac{1}{2} \text{Re}\big[ {T}^{Z,ab}_{n} + {T}^{Z,ba}_{n} \big]
    = \frac{1}{4\mathcal{E}} \left[ (\hat{\vect{d}} \times \vect{v}_a)_b + (\hat{\vect{d}} \times \vect{v}_b)_a \right], \label{supp:eq:gN} \\
    g^{A,ab}_{n} &\equiv \frac{1}{2} \text{Im}\big[ {T}^{Z,ab}_{n} + {T}^{Z,ba}_{n} \big]
    = -\frac{1}{4n\mathcal{E}} \left[ (\vect{v}_{a,\perp})_b + (\vect{v}_{b,\perp})_a \right], \label{supp:eq:gA} \\
    \Omega^{N,ab}_{n} &\equiv -\text{Im}\big[ {T}^{Z,ab}_{n} - {T}^{Z,ba}_{n} \big]
    = \frac{1}{2n\mathcal{E}} \left[ (\vect{v}_{a,\perp})_b - (\vect{v}_{b,\perp})_a \right], \label{supp:eq:OmegaN} \\
    \Omega^{A,ab}_{n} &\equiv -\text{Re}\big[ {T}^{Z,ab}_{n} - {T}^{Z,ba}_{n} \big]
    = -\frac{1}{2\mathcal{E}} \left[ (\hat{\vect{d}} \times \vect{v}_a)_b - (\hat{\vect{d}} \times \vect{v}_b)_a \right]. \label{supp:eq:OmegaA}
\end{align}
Equations \eqref{supp:eq:gN}--\eqref{supp:eq:OmegaA} are \emph{universal} for any two-band Hamiltonian \(H = \vect{d} \cdot \bm{\sigma}\).

\subsection{Simplification for \(\Omega^{N}\) and \(\Omega^{A}\)}
For a two-band Hamiltonian \(H = \vect{d} \cdot \bm{\sigma}\) with \(\bd = \vect{d}/\mathcal{E}\), we define \(\vect{v}_a = \partial_{k_a} \vect{d}\) for \(a \in \{x, y, z\}\). The universal two-band identities for the normal and anomalous Zeeman Berry curvatures read
\begin{equation}
    \Omega^{N,ab}_{n} = \frac{(\vect{v}_{a,\perp})_b - (\vect{v}_{b,\perp})_a}{2n\mathcal{E}}, \qquad
    \Omega^{A,ab}_{n} = -\frac{(\hat{\vect{d}} \times \vect{v}_a)_b - (\hat{\vect{d}} \times \vect{v}_b)_a}{2\mathcal{E}},
    \label{supp:eq:start}
\end{equation}
where \(n = \pm 1\) labels the bands (\(+\) or \(-\)), and \(\vect{v}_{a,\perp} \equiv \vect{v}_a - \bd (\bd \cdot \vect{v}_a)\) is the component transverse to \(\bd\). Now we reduce these to compact vector (axial) forms built solely from \(\bd(\vect{k})\).

Here we derive the identities
\begin{equation}
    \vect{v}_{a,\perp} = \mathcal{E} \partial_{k_a} \hat{\vect{d}}, \qquad \hat{\vect{d}} \times \vect{v}_a = \mathcal{E} (\hat{\vect{d}} \times \partial_{k_a} \hat{\vect{d}}). \label{supp:eq:C1}
\end{equation}

Starting from \(\vect{d} = \mathcal{E}\hat{\vect{d}}\), differentiation gives
\begin{equation}
    \partial_{k_a} \vect{d} = \partial_{k_a}\mathcal{E}\, \hat{\vect{d}} + \mathcal{E} \partial_{k_a} \hat{\vect{d}}. \label{supp:eq:C2}
\end{equation}
Thus,
\begin{equation}
    \vect{v}_a = \partial_{k_a}\mathcal{E}\, \hat{\vect{d}} + \mathcal{E} \partial_{k_a} \hat{\vect{d}}. \label{supp:eq:C3}
\end{equation}

\noindent{1. Calculation of \(\vect{v}_{a,\perp} = \mathcal{E} \partial_{k_a} \hat{\vect{d}}\).}

First, the transverse component of \(\vect{v}_{a}\) is defined as
\begin{equation}
    \vect{v}_{a,\perp} = \vect{v}_a - \hat{\vect{d}} (\hat{\vect{d}} \cdot \vect{v}_a). \label{supp:eq:C4}
\end{equation}
Substituting Eq.~\eqref{supp:eq:C3}, we find
\begin{equation}
    \vect{v}_{a,\perp} = \big[ \partial_{k_a}\mathcal{E}\, \hat{\vect{d}} + \mathcal{E} \partial_{k_a} \hat{\vect{d}} \big] - \hat{\vect{d}} \big( \hat{\vect{d}} \cdot \big[ \partial_{k_a}\mathcal{E}\, \hat{\vect{d}} + \mathcal{E} \partial_{k_a} \hat{\vect{d}} \big] \big). \label{supp:eq:C5}
\end{equation}
The dot product in Eq.~\eqref{supp:eq:C4} is calculated as
\begin{equation}
    \hat{\vect{d}} \cdot \vect{v}_a = \hat{\vect{d}} \cdot \big[ \partial_{k_a}\mathcal{E}\, \hat{\vect{d}} + \mathcal{E} \partial_{k_a} \hat{\vect{d}} \big] = \partial_{k_a}\mathcal{E}\,(\hat{\vect{d}} \cdot \hat{\vect{d}}) + \mathcal{E}(\hat{\vect{d}} \cdot \partial_{k_a} \hat{\vect{d}}). \label{supp:eq:C6}
\end{equation}
Since \(\hat{\vect{d}}\) is a unit vector, \(\hat{\vect{d}} \cdot \hat{\vect{d}} = 1\), and \(\hat{\vect{d}} \cdot \partial_{k_a} \hat{\vect{d}} = 0\), which follow from \(0 = \partial_{k_a} (\hat{\vect{d}} \cdot \hat{\vect{d}}) = 2 \hat{\vect{d}} \cdot \partial_{k_a} \hat{\vect{d}}\). Therefore,
\begin{equation}
    \hat{\vect{d}} \cdot \vect{v}_a = \partial_{k_a}\mathcal{E}. \label{supp:eq:C7}
\end{equation}
Substituting this result into Eq.~\eqref{supp:eq:C4} gives
\begin{equation}
    \vect{v}_{a,\perp} = \big[ \partial_{k_a}\mathcal{E}\, \hat{\vect{d}} + \mathcal{E} \partial_{k_a} \hat{\vect{d}} \big] - \hat{\vect{d}}\,\partial_{k_a}\mathcal{E} = \mathcal{E} \partial_{k_a} \hat{\vect{d}}. \label{supp:eq:C8}
\end{equation}

\noindent{2. Calculation of \(\hat{\vect{d}} \times \vect{v}_a = \mathcal{E}(\hat{\vect{d}} \times \partial_{k_a} \hat{\vect{d}})\).}

Second, using Eq.~\eqref{supp:eq:C3}, we obtain
\begin{equation}
\hat{\vect{d}} \times \vect{v}_a
= \hat{\vect{d}} \times \big[ \partial_{k_a}\mathcal{E}\, \hat{\vect{d}} + \mathcal{E} \partial_{k_a} \hat{\vect{d}} \big]. \label{supp:eq:C9}
\end{equation}
Since \(\hat{\vect{d}} \times \hat{\vect{d}} = 0\), the first term vanishes:
\begin{equation}
    \hat{\vect{d}} \times [\partial_{k_a}\mathcal{E}\, \hat{\vect{d}}]
    = \partial_{k_a}\mathcal{E}\,(\hat{\vect{d}} \times \hat{\vect{d}}) = 0 . \label{supp:eq:C10}
\end{equation}
Therefore,
\begin{equation}
    \hat{\vect{d}} \times \vect{v}_a
    = \hat{\vect{d}} \times [\mathcal{E} \partial_{k_a} \hat{\vect{d}}] = \mathcal{E}(\hat{\vect{d}} \times \partial_{k_a} \hat{\vect{d}}). \label{supp:eq:C11}
\end{equation}

Substitute Eq.~\eqref{supp:eq:C8} and Eq.~\eqref{supp:eq:C11} into Eq.~\eqref{supp:eq:start}, the explicit \(\mathcal{E}\) cancels
\begin{align}
    \Omega^{N,ab}_{n} &= \frac{1}{2n} \left[ \partial_{k_a} \hat{d}_b - \partial_{k_b} \hat{d}_a \right], \label{supp:eq:OmegaN-hatd} \\
    \Omega^{A,ab}_{n} &= -\frac{1}{2} \left[ (\bd \times \partial_{k_a} \bd)_b - (\bd \times \partial_{k_b} \bd)_a \right]. \label{supp:eq:OmegaA-hatd}
\end{align}

Starting from Eq.~\eqref{supp:eq:OmegaN-hatd},
we employ the Levi--Civita duality between antisymmetric tensors and vectors
\begin{equation}
    (\Omega^{N}_{n})_c = \frac{1}{2} \eps_{cab} \, \Omega^{N,ab}_{n}
    \qquad \Longleftrightarrow \qquad
    \Omega^{N,ab}_{n} = \eps_{abc} \, (\Omega^{N}_{n})_c,
    \label{supp:eq:duality}
\end{equation}
where $\eps_{abc}$ is the Levi-Civita symbol, and we obtain
\begin{align}
    (\Omega^{N}_{n})_c = \frac{1}{4n} \, \eps_{cab} \left[ \partial_{k_a} \hat{d}_b - \partial_{k_b} \hat{d}_a \right].
\end{align}
Relabeling \(a \leftrightarrow b\) in the second term and using the antisymmetry of \(\eps_{cab}\), one finds that the two terms are equal. Hence
\begin{align}
    (\Omega^{N}_{n})_c = \frac{1}{2n} \, \eps_{cab} \, \partial_{k_a} \hat{d}_b.
\end{align}
Using \((\vk \times \bd)_c = \eps_{cab} \, \partial_{k_a} \hat{d}_b\) and \(1/n = n\), the \emph{vector form} of \(\Omega^{N}_{n}\) is
\begin{equation}
        \bm{\Omega}^{N}_{n} = \frac{n}{2} \, \vk \times \bd .
    \label{supp:eq:OmegaN-vector-prelim}
\end{equation}

Similarly, we have
\begin{align}
    (\Omega^{A}_{n})_c=  -\frac{1}{4} \eps_{cab} \left[ (\bd \times \partial_{k_a} \bd)_b - (\bd \times \partial_{k_b} \bd)_a \right].
\end{align}
Relabeling \(a \leftrightarrow b\) in the second terms gives
\begin{equation}
    (\Omega^{A}_{n})_c = -\frac{1}{2} \eps_{cab} \, (\bd \times \partial_{k_a} \bd)_b.
    \label{supp:eq:OmegaA-mid}
\end{equation}
Using the identity \(\eps_{cab}(\hat{\vect{d}} \times \partial_{k_a}\hat{\vect{d}})_b = \hat{d}_c(\nabla_k\cdot\hat{\vect{d}}) - (\hat{\vect{d}}\cdot\nabla_k)\hat{d}_c\), we obtain the vector form
\begin{equation}\label{supp:eq:OmegaA-vector-final}
    \bm{\Omega}^{A}_{n} = -\frac{1}{2}\bigl[ \bd(\vk\cdot\bd) - (\bd\cdot\vk)\bd \bigr].
\end{equation}

\subsection{Simplification for \(g^{N}\) and \(g^{A}\)}
For the two-band Hamiltonian \(H = \vect{d} \cdot \bm{\sigma}\), the universal decomposition of \({T}^{Z,ab}_{n}\) gives the following \emph{symmetric} tensors (in \(a \leftrightarrow b\)):
\begin{equation}
g^{N,ab}_{n} = \frac{(\bd \times \vect{v}_a)_b + (\bd \times \vect{v}_b)_a}{4\mathcal{E}}, \qquad
g^{A,ab}_{n} = -\frac{(\vect{v}_{a,\perp})_b + (\vect{v}_{b,\perp})_a}{4n\mathcal{E}}. \label{supp:eq:start-sym}
\end{equation}
where \(n = \pm 1\) labels the bands (\(+\) or \(-\)).

Substituting Eq.~\eqref{supp:eq:C8} and Eq.~\eqref{supp:eq:C11} into Eq.~\eqref{supp:eq:start-sym}, the explicit \(\mathcal{E}\) cancels, which gives
\begin{align}
g^{N,ab}_{n} &= \frac{1}{4} \left[ (\bd \times \partial_{k_a} \bd)_b + (\bd \times \partial_{k_b} \bd)_a \right], \label{supp:eq:gN-hatd} \\
g^{A,ab}_{n} &= -\frac{n}{4} \left[ \partial_{k_a} \hat{d}_b + \partial_{k_b} \hat{d}_a \right]. \label{supp:eq:gA-hatd}
\end{align}

\subsection{2D Dirac semimetal}
For a planar texture \(\hat{\vect{d}} = (\cos\theta_{\vect{k}}, \sin\theta_{\vect{k}}, 0)\), we define the winding field as \(\bm{w} = \nabla_{\vect{k}} \theta_{\vect{k}}\). For \(\theta_{\vect{k}} = \arctan(k_y/k_x)\), the gradient is
\begin{equation}
    \nabla_{\vect{k}} \theta_{\vect{k}} = (\partial_{k_x} \theta_{\vect{k}}, \partial_{k_y} \theta_{\vect{k}}).
\end{equation}
A direct calculation gives
\begin{align}
    \partial_{k_x} \theta_{\vect{k}} &= \frac{1}{1 + (k_y/k_x)^2} \cdot \left( -\frac{k_y}{k_x^2} \right) = -\frac{k_y}{k_x^2 + k_y^2}, \\
    \partial_{k_y} \theta_{\vect{k}} &= \frac{1}{1 + (k_y/k_x)^2} \cdot \frac{1}{k_x} = \frac{k_x}{k_x^2 + k_y^2}.
\end{align}
Therefore,
\begin{equation}
    \nabla_{\vect{k}} \theta_{\vect{k}} = \bigl( -k_y, \, k_x \bigr) / k^2, \qquad k^2 = k_x^2 + k_y^2.
\end{equation}

Using the Hodge star convention \(* (v_x, v_y) = (-v_y, v_x)\), we find
\begin{equation}
    * \nabla_{\vect{k}} \theta_{\vect{k}} = * \bigl( -k_y, \, k_x \bigr) / k^2 = -\bigl( k_x, \, k_y \bigr) / k^2 = -\frac{\vect{k}}{k^2}.
\end{equation}

Substituting \(\hat{\vect{d}} = (\cos\theta_{\vect{k}}, \sin\theta_{\vect{k}}, 0)\) into Eq.~\eqref{supp:eq:OmegaA-vector-final}, we obtain
\begin{equation}
    \bm{\Omega}^A = -\frac{1}{2} \frac{\vect{k}}{k^2} = \frac{* \nabla_{\vect{k}} \theta_{\vect{k}}}{2} = \frac{1}{2} \bigl( -\partial_{k_y} \theta_{\vect{k}}, \, \partial_{k_x} \theta_{\vect{k}} \bigr).
\end{equation}

We now compute the \(z\)-component of \(\nabla_{\vect{k}} \times \bm{\Omega}^A\)
\begin{align}
    [\nabla_{\vect{k}} \times \bm{\Omega}^A]_z
    &= \partial_{k_x} \Omega_y^A - \partial_{k_y} \Omega_x^A
    = \frac{1}{2} \bigl( \partial_{k_x}^2 \theta_{\vect{k}} + \partial_{k_y}^2 \theta_{\vect{k}} \bigr) \nonumber \\
    &= \frac{1}{2} \nabla_{\vect{k}} \cdot \bm{w}.
\end{align}
Similarly, \(\nabla_{\vect{k}} \cdot \bm{\Omega}^A\) is
\begin{align}
    \nabla_{\vect{k}} \cdot \bm{\Omega}^A
    &= \partial_{k_x} \Omega_x^A + \partial_{k_y} \Omega_y^A
    = -\frac{1}{2} \bigl( \partial_{k_x} \partial_{k_y} \theta_{\vect{k}} - \partial_{k_y} \partial_{k_x} \theta_{\vect{k}} \bigr) \nonumber \\
    &= -\frac{1}{2} [\nabla_{\vect{k}} \times \bm{w}]_z.
\end{align}

\phantomsection
\label{app:linear-response}
\section*{Appendix B: Linear response theory for the current density and spin magnetization}

\subsection{Liouville equation and expansion of the density matrix}
Consider the system Hamiltonian \(H = H_0 + H_1\), where \(H_0\) describes the periodic crystal and \(H_1\) denotes the perturbation due to external fields. The perturbing Hamiltonian is given by (\(e = \hbar = 1\))
\begin{equation}
    H_1 = E^\alpha(t) r^\alpha - g \mu_B B^\alpha(t) \sigma^\alpha,
\end{equation}
with
\begin{align}
    E^\alpha(t) &= \frac{1}{2} E^\alpha \sum_{\omega_\beta = \pm \omega_0} e^{-i\omega_\beta t}, \\
    B^\alpha(t) &= \frac{1}{2} B^\alpha \sum_{\omega_\beta = \pm \omega_0} e^{-i\omega_\beta t},
\end{align}
being the external fields oscillating at a single frequency.

The density matrix \(\rho\) of the system satisfies the quantum Liouville equation~\cite{Aversa1995}
\begin{equation}
    i \partial_t \rho = [H, \rho].
\end{equation}

In the Bloch basis, the density matrix elements \(\rho_{mn}\) satisfy
\begin{equation}
    i \partial_t \rho_{mn} = \epsilon_{mn} \rho_{mn} + i (\rho_{mn})_{;k_\alpha} E^\alpha(t) + \sum_l \left[ r_{ml}^\alpha \rho_{ln} - \rho_{ml} r_{ln}^\alpha \right] E^\alpha(t) - g \mu_B B^\alpha(t) \sum_l \left[ \sigma_{ml}^\alpha \rho_{ln} - \rho_{ml} \sigma_{ln}^\alpha \right],
\end{equation}
where \(\epsilon_{mn} = \epsilon_m - \epsilon_n\) is the energy difference. \((\rho_{mn})_{;k_\alpha} = \left[ \partial_{k_\alpha} - i (\mathcal{A}_m^\alpha - \mathcal{A}_n^\alpha) \right] \rho_{mn}\) is the covariant derivative with \(\mathcal{A}_m^\alpha\) the intraband Berry connection. \(r_{ml}^\alpha\) is the interband Berry connection. \(\sigma_{ml}^\alpha\) is the matrix element of the Pauli matrices.

The density matrix is expanded in powers of the external fields as \(\rho = \rho^{(0)} + \rho^{(1)} + \cdots\), with \(\rho_{mn}^{(0)} = \delta_{mn} f_n\), where \(f_n = f(\epsilon_n)\) is the Fermi--Dirac distribution function. Substituting this expansion into the Liouville equation and retaining terms linear in the external fields gives the equation for the first-order density matrix:
\begin{equation}
    i \partial_t \rho_{mn}^{(1)} = \epsilon_{mn} \rho_{mn}^{(1)} + i \delta_{mn} \partial_\alpha f_m E^\alpha(t) + f_{nm} r_{mn}^\alpha E^\alpha(t) - g \mu_B B^\alpha(t) f_{nm} \sigma_{mn}^\alpha,
\end{equation}
with \(f_{nm} = f_n - f_m\). Solving this equation, a steady-state solution of the form \(\rho_{mn}^{(1)} \propto e^{-i\omega_\beta t}\) yields
\begin{equation}\label{supp:eq:density}
    \rho_{mn}^{(1)} = \frac{1}{2} \sum_{\omega_\beta = \pm \omega_0} \left[ i \delta_{mn} \partial_\alpha f_m + f_{nm} r_{mn}^\alpha \right] \frac{E^\alpha e^{-i\omega_\beta t}}{\omega_\beta - \epsilon_{mn} + i\eta} - \frac{g \mu_B}{2} \sum_{\omega_\beta = \pm \omega_0} \frac{f_{nm} \sigma_{mn}^\alpha B^\alpha e^{-i\omega_\beta t}}{\omega_\beta - \epsilon_{mn} + i\eta}.
\end{equation}
Here \(\eta\) is a positive infinitesimal ensuring causality.

\subsection{Derivation of Eq.~(\ref{exp}) in the main text}
The linear current density is given by
\begin{equation}\label{supp:eq:current}
    J^a(t) = \sum_{nm} \int_k v^a_{nm} \rho^{(1)}_{mn}(t),
\end{equation}
where \(\int_k = \int \frac{\dd^d k}{(2\pi)^d}\).

Substituting the magnetic field component of Eq.~\eqref{supp:eq:density} into Eq.~\eqref{supp:eq:current}, the linear current density driven by a time-dependent magnetic field is given by
\begin{equation}
    J^{a} = - \frac{g \mu_B}{2} \sum_{nm,\beta} \int_{k} \frac{v_{nm}^{a} \sigma_{mn}^{b} f_{nm}}{\omega_{\beta} - \epsilon_{mn} + i\eta} B^{b} e^{-i\omega_{\beta}t},
\end{equation}
where \(v_{nm}^{a} = i \epsilon_{nm} r_{nm}^{a}\).

For the non-resonant part (\(\omega_{\beta}\) far from \(\epsilon_{mn}\)), the imaginary part \(i\eta\) can be neglected, and the principal value integral is used. This gives
\begin{equation}
    J^{a} = - \frac{g \mu_B}{2} \sum_{nm,\beta} \int_{k} \frac{v_{nm}^{a} \sigma_{mn}^{b} f_{nm}}{\omega_{\beta} - \epsilon_{mn}} B^{b} e^{-i\omega_{\beta}t}.
\end{equation}

The current density is related to the conductivity tensor \(\sigma^{ab}_\beta\) via \(J^{a}_\beta = \sigma^{ab}_\beta B^{b} e^{-i\omega_{\beta}t}\), so that
\begin{align}
    \sigma^{ab}_\beta = - \frac{g \mu_B}{2} \sum_{nm} \int_{k} \frac{i \epsilon_{nm} r_{nm}^{a} \sigma_{mn}^{b} f_{nm} (\omega_{\beta} + \epsilon_{mn})}{\omega_{\beta}^{2} - \epsilon_{mn}^{2}}.
\end{align}

Substituting \(f_{nm} = f_n - f_m\) and using the relation \(\sum_{nm} f_{nm} A_{nm} = \sum_{nm} f_n (A_{nm} - A_{mn})\), we obtain
\begin{align}
    \sigma^{ab}_\beta = - \frac{g \mu_B}{2} \sum_{nm} \int_{k} \frac{i f_{n} \epsilon_{nm}}{\omega_{\beta}^{2} - \epsilon_{mn}^{2}} \Bigl[ \omega_{\beta} \bigl( r_{nm}^{a} \sigma_{mn}^{b} + r_{mn}^{a} \sigma_{nm}^{b} \bigr) + \epsilon_{mn} \bigl( r_{nm}^{a} \sigma_{mn}^{b} - r_{mn}^{a} \sigma_{nm}^{b} \bigr) \Bigr].
\end{align}

Since
\begin{align}
    \text{Re}(r_{nm}^{a} \sigma_{mn}^{b}) &= \frac{1}{2} \bigl( r_{nm}^{a} \sigma_{mn}^{b} + r_{mn}^{a} \sigma_{nm}^{b} \bigr), \\
    \text{Im}(r_{nm}^{a} \sigma_{mn}^{b}) &= -\frac{i}{2} \bigl( r_{nm}^{a} \sigma_{mn}^{b} - r_{mn}^{a} \sigma_{nm}^{b} \bigr),
\end{align}
we have
\begin{equation}\label{supp:eq:sigma}
    \sigma^{ab}_\beta = -g \mu_B \sum_{nm} \int_{k} \frac{f_{n} \epsilon_{nm}}{\omega_{\beta}^{2} - \epsilon_{mn}^{2}} \Bigl[ i \omega_{\beta} \text{Re}(r_{nm}^{a} \sigma_{mn}^{b}) - \epsilon_{mn} \text{Im}(r_{nm}^{a} \sigma_{mn}^{b}) \Bigr].
\end{equation}

Under the low-frequency condition (\(\omega_{\beta} \ll \epsilon_{nm}\)), we expand the denominator and retain terms up to \(\omega_\beta^2\), which gives
\begin{equation}
    \sigma^{ab}_\beta \approx g\mu_B \sum_{nm}\int_k f_n\left[ i\frac{\omega_\beta}{\epsilon_{nm}}\mathrm{Re}(r_{nm}^a\sigma_{mn}^b) + \left(1+\frac{\omega_\beta^2}{\epsilon_{nm}^2}\right)\mathrm{Im}(r_{nm}^a\sigma_{mn}^b) \right].
\end{equation}
Then we obtain
\begin{align}
    (\mathbf{B} \times \bm{\Omega}_{nm}^{A})^a= -\varepsilon_{abc} B^b \text{Re}( \varepsilon_{cde} r_{nm}^{d} \sigma_{mn}^{e} ) = -\text{Re}\big[ r_{nm}^{a} \sigma_{mn}^{b} - r_{nm}^{b} \sigma_{mn}^{a} \big] B^b,
\end{align}
and
\begin{equation}
    \mathbf{g}_{nm}^{N} \cdot \mathbf{B} = \mathbf{g}^{N,ab}_{nm} B^b = \frac{1}{2} \text{Re}\big( r_{nm}^{a} \sigma_{mn}^{b} + r_{nm}^{b} \sigma_{mn}^{a} \big) B^b.
\end{equation}

Similarly,
\begin{align}
    (\mathbf{B} \times \bm{\Omega}_{nm}^{N})^a &= -\text{Im}\big( r_{nm}^{a} \sigma_{mn}^{b} - r_{nm}^{b} \sigma_{mn}^{a} \big) B^b, \\
    \mathbf{g}_{nm}^{A} \cdot \mathbf{B} &= \frac{1}{2} \text{Im}\big( r_{nm}^{a} \sigma_{mn}^{b} + r_{nm}^{b} \sigma_{mn}^{a} \big) B^b.
\end{align}

Then, at low frequencies, the current is expressed as
\begin{equation}
    \mathbf{J}_\beta = -\frac{g \mu_B}{2} \sum_{nm} \int_{k} f_{n} \left[ i \frac{\omega_{\beta}}{\epsilon_{nm}} \left( \mathbf{B} \times \bm{\Omega}_{nm}^{A} - 2 \mathbf{g}_{nm}^{N} \cdot \mathbf{B} \right) + \left( 1 + \frac{\omega_{\beta}^{2}}{\epsilon_{nm}^{2}} \right) \left( \mathbf{B} \times \bm{\Omega}_{nm}^{N} - 2 \mathbf{g}_{nm}^{A} \cdot \mathbf{B} \right) \right] e^{-i \omega_\beta t}.
\end{equation}

Using the Fourier transform \(\mathcal{F}[e^{-i \omega_\beta t}] = \delta(\omega + \omega_\beta)\), we have
\begin{equation}
\begin{aligned}
    \mathbf{J}(\omega) = -\frac{g \mu_B}{2} \sum_{nm,\beta} \int_{k} f_{n} \Bigg[ \frac{i \omega_\beta}{\epsilon_{nm}} \left( \mathbf{B} \times \bm{\Omega}_{nm}^{A} - 2 \mathbf{g}_{nm}^{N} \cdot \mathbf{B} \right) + \left( 1 + \frac{\omega_\beta^{2}}{\epsilon_{nm}^{2}} \right) \left( \mathbf{B} \times \bm{\Omega}_{nm}^{N} - 2 \mathbf{g}_{nm}^{A} \cdot \mathbf{B} \right) \Bigg] \delta(\omega + \omega_\beta).
\end{aligned}
\end{equation}

Summing over $\beta$, we obtain
\begin{equation}
\begin{aligned}
\mathbf{J}(\omega) = -{g \mu_B} \sum_{nm} \int_{k} f_{n} \Bigg[ \frac{i \omega}{\epsilon_{nm}} \left( 2 \mathbf{g}_{nm}^{N} \cdot \mathbf{B}(\omega)-\mathbf{B}(\omega) \times \bm{\Omega}_{nm}^{A} \right)
    + \left( 1 + \frac{\omega^{2}}{\epsilon_{nm}^{2}} \right) \left( \mathbf{B}(\omega) \times \bm{\Omega}_{nm}^{N} - 2 \mathbf{g}_{nm}^{A} \cdot \mathbf{B}(\omega) \right) \Bigg].
    \end{aligned}
\end{equation}
where $\mathbf{B}(\omega)=\frac12\mathbf{B}[\delta(\omega + \omega_0)+\delta(\omega - \omega_0)]$.

\subsection{Derivation of Eq.~(\ref{exp1}) in the main text}
From Eq.~\eqref{supp:eq:density}, considering only the electric field contribution, we have
\begin{equation}
    \rho_{mn}^{(1)} = \frac{1}{2} \sum_{\omega_\beta = \pm \omega_0} \left[ i \delta_{mn} \partial_\alpha f_m + f_{nm} r_{mn}^\alpha \right] \frac{E^\alpha e^{-i\omega_\beta t}}{\omega_\beta - \epsilon_{mn} + i\eta}.
\end{equation}
For insulators or semiconductors, we can focus solely on the contribution from the interband terms, which gives
\begin{equation}
    \rho_{mn}^{(1)} = \frac{1}{2} \sum_{\omega_\beta = \pm \omega_0} \frac{f_{nm} r_{mn}^\alpha E^\alpha e^{-i\omega_\beta t}}{\omega_\beta - \epsilon_{mn} + i\eta}. \label{supp:Eq51}
\end{equation}

The spin magnetization induced by a time-dependent electric field is defined as
\begin{equation}
    \mathbf{M} \equiv \text{Tr}[\hat{\bm{s}} \hat{\rho}^{(1)}] = \frac{1}{2} \sum_{nm} \int_k \bm{\sigma}_{nm} \rho^{(1)}_{mn},
\end{equation}
where \(\hat{\bm{s}}\) is the spin operator and \(\hat{\bm{s}} = \frac{1}{2} \hat{\bm{\sigma}}\).

Substituting \(\rho_{nm}^{(1)}\) from Eq.~\eqref{supp:Eq51} and focusing on the non-resonant part (ignoring \(i\eta\)), we have
\begin{align}
    \mathbf{M} = \frac{1}{4} \sum_{nm, \beta} \int_k \frac{ \omega_\beta \sigma^a_{nm} r^b_{mn} + \epsilon_{mn} \sigma^a_{nm} r^b_{mn} }{ \omega_\beta^2 - \epsilon_{mn}^2 } f_{nm} E^b e^{-i\omega_\beta t}.
\end{align}
Using \(f_{nm} = f_n - f_m\), we can rewrite the summation as
\begin{equation}
    M^a = \frac{1}{4} \sum_{nm, \beta} \int_k f_n \frac{ \omega_\beta (\sigma^a_{nm} r^b_{mn} - \sigma^a_{mn} r^b_{nm}) + \epsilon_{mn} (\sigma^a_{nm} r^b_{mn} + \sigma^a_{mn} r^b_{nm}) }{ \omega_\beta^2 - \epsilon_{mn}^2 } E^b e^{-i\omega_\beta t}.
\end{equation}

Noting that \(\text{Re}(\sigma_{nm}^{a} r_{mn}^{b}) = \frac{1}{2} (\sigma_{nm}^{a} r_{mn}^{b} + \sigma_{mn}^{a} r_{nm}^{b})\) and \(\text{Im}(\sigma_{nm}^{a} r_{mn}^{b}) = -\frac{i}{2} (\sigma_{nm}^{a} r_{mn}^{b} - \sigma_{mn}^{a} r_{nm}^{b})\), substitution gives
\begin{equation}
    M^a = \frac{1}{2} \sum_{nm, \beta} \int_k f_n \frac{ i \omega_\beta \text{Im}(\sigma_{nm}^{a} r_{mn}^{b}) + \epsilon_{mn} \text{Re}(\sigma_{nm}^{a} r_{mn}^{b}) }{ \omega_\beta^2 - \epsilon_{mn}^2 } E^b e^{-i\omega_\beta t}.
\end{equation}
Under the low-frequency condition (\(\omega_{\beta} \ll \epsilon_{nm}\)), keeping terms up to order \(\omega_\beta^2\) and neglecting higher orders, we have
\begin{equation}\label{supp:eq:sa}
    M^a \approx -\frac{1}{2} \sum_{nm, \beta} \int_k f_n \left[ \frac{i \omega_\beta}{\epsilon_{mn}^2} \text{Im}(\sigma_{nm}^{a} r_{mn}^{b}) + \left( \frac{1}{\epsilon_{mn}} + \frac{\omega_\beta^2}{\epsilon_{mn}^3} \right) \text{Re}(\sigma_{nm}^{a} r_{mn}^{b}) \right] E^b e^{-i\omega_\beta t}.
\end{equation}
Using the relations $\text{Im}(\sigma_{nm}^{a} r_{mn}^{b})=-\text{Im}(\sigma_{mn}^{a} r_{nm}^{b})$ and $\text{Re}(\sigma_{nm}^{a} r_{mn}^{b})=\text{Re}(\sigma_{mn}^{a} r_{nm}^{b})$, we can rewrite $S^a$ as
\begin{equation}
    M^a = \frac{1}{2} \sum_{nm, \beta} \int_k f_n \left[ \frac{i \omega_\beta}{\epsilon_{nm}^2} \text{Im}(\sigma_{mn}^{a} r_{nm}^{b}) + \left( \frac{1}{\epsilon_{nm}} + \frac{\omega_\beta^2}{\epsilon_{nm}^3} \right) \text{Re}(\sigma_{mn}^{a} r_{nm}^{b}) \right] E^b e^{-i\omega_\beta t}.
\end{equation}

Using the following identities
\begin{align}
    (\mathbf{E} \times \bm{\Omega}_{nm}^{A})^a &= -\text{Re}\big[ r_{nm}^{a} \sigma_{mn}^{b} - r_{nm}^{b} \sigma_{mn}^{a} \big] E^b, \\
    \mathbf{g}_{nm}^{N} \cdot \mathbf{E} &= \frac{1}{2} \text{Re}\big( r_{nm}^{a} \sigma_{mn}^{b} + r_{nm}^{b} \sigma_{mn}^{a} \big) E^b, \\
    (\mathbf{E} \times \bm{\Omega}_{nm}^{N})^a &= -\text{Im}\big( r_{nm}^{a} \sigma_{mn}^{b} - r_{nm}^{b} \sigma_{mn}^{a} \big) E^b, \\
    \mathbf{g}_{nm}^{A} \cdot \mathbf{E} &= \frac{1}{2} \text{Im}\big( r_{nm}^{a} \sigma_{mn}^{b} + r_{nm}^{b} \sigma_{mn}^{a} \big) E^b,
\end{align}
we obtain
\begin{equation}
    \mathbf{M} = \sum_{nm, \beta} \int_k \frac{f_n}{4 \epsilon_{nm}^2} \Bigg[ i \omega_\beta \left( \mathbf{E} \times \bm{\Omega}^N_{nm} + 2 \mathbf{g}^A_{nm} \cdot \mathbf{E} \right) + \left( \epsilon_{nm} + \frac{\omega_\beta^2}{\epsilon_{nm}} \right) \left( \mathbf{E} \times \bm{\Omega}^A_{nm} + 2 \mathbf{g}^N_{nm} \cdot \mathbf{E} \right) \Bigg] e^{-i\omega_\beta t}.
\end{equation}
By Fourier transform \(\mathcal{F}[e^{-i \omega_\beta t}] = \delta(\omega + \omega_\beta)\) and summing over $\beta$, we finally have
\begin{equation}
    \mathbf{M}(\omega) = \sum_{nm} \int_k \frac{f_n}{2 \epsilon_{nm}^2} \Bigg[ -i \omega \left( \mathbf{E}(\omega) \times \bm{\Omega}^N_{nm} + 2 \mathbf{g}^A_{nm} \cdot \mathbf{E}(\omega) \right) + \left( \epsilon_{nm} + \frac{\omega^2}{\epsilon_{nm}} \right) \left( \mathbf{E}(\omega) \times \bm{\Omega}^A_{nm} + 2 \mathbf{g}^N_{nm} \cdot \mathbf{E}(\omega) \right) \Bigg].
\end{equation}
Here $\mathbf{E}(\omega)=\frac12\mathbf{E}[\delta(\omega + \omega_0)+\delta(\omega - \omega_0)]$.

\phantomsection
\label{app:spin-current}
\section*{Appendix C: Linear spin current density in terms of the sectors of spin QGT}

In this appendix, we derive the electric-field-induced spin current and show that it admits a natural decomposition in terms of the four sectors of the spin quantum geometric tensor.

The intrinsic spin Hall conductivity is\cite{universal} ($\hbar=e=1$)
\begin{equation}
\sigma_{ab,c}^{s}
=
-2\int_{\bm{k}} \sum_{nm}
f_n\,
\mathrm{Im}
\frac{v_{nm}^{a,c} v_{mn}^{b}}{\epsilon_{nm}^{2}},
\label{eq:B1}
\end{equation}

Using the definitions of the spin quantum geometric tensor, Eq.~(\ref{eq:B1}) can be rewritten as
\begin{equation}
\sigma_{ab,c}^{s}
=
-\int_{\bm{k}} \sum_{nm} f_n
\left[
2(g_s^{A})_{nm}^{ab,c}
-
(\Omega_s^{N})_{nm}^{ab,c}
\right].
\label{eq:B2}
\end{equation}

We now consider the nonequilibrium contribution induced by a dc electric field \(\mathbf{E}\). Within the relaxation-time approximation, the distribution function acquires the correction
\begin{equation}
\delta f_n=\tau E_b \partial_{k_b} f_n,
\end{equation}
which generates a spin current
\begin{equation}
J_n^{a,c}
=
\tau \int_{\bm{k}} v_{nn}^{a,c}\,\partial_{k_b}f_n\,E_b.
\end{equation}
Integrating by parts gives
\begin{equation}
J_n^{a,c}
=
-\tau \int_{\bm{k}} f_n\,\partial_{k_b}v_{nn}^{a,c}\,E_b.
\label{eq:B3}
\end{equation}

To evaluate \(\partial_{k_b}v_{nn}^{a,c}\), we use
\begin{equation}
\partial_{k_b} v_{nn}^{a,c}
=
i\sum_m
\left(
r_{nm}^{b}v_{mn}^{a,c}
-
v_{nm}^{a,c}r_{mn}^{b}
\right)
+
\frac{1}{2}
\left\{
\partial_{k_b}v^{a},\sigma^{c}
\right\}_{nn}.
\label{eq:B4}
\end{equation}
Using \(v_{nm}^{b}=i\epsilon_{nm}r_{nm}^{b}\), the interband part can be expressed directly in terms of the spin geometric tensors. Neglecting the last term in Eq.~(\ref{eq:B4}), which vanishes for Hamiltonians linear in momentum, we obtain
\begin{equation}
\bm{J}_{\mathrm{ext}}^{\,c}
=
-\tau \int_{\bm{k}} \sum_n f_n
\left[
2\,\bar g_n^{N,c}\cdot \mathbf{E}
-
\mathbf{E}\times \bar{\bm{\Omega}}_n^{A,c}
\right],
\label{eq:B5}
\end{equation}
where $\bar{\bm{\Omega}}_n^{A,c}$ and $\bar g_n^{N,c}$ are defined in Eq.~(\ref{bar}).

Combining the intrinsic and nonequilibrium (extrinsic) contributions, the total spin current takes the compact form
\begin{equation}
\bm{J}^{\,c} = \int_{\bm{k}} \sum_n f_n \left[ \mathbf{E}\times \left( {\bm{\Omega}}_n^{N,c} + \tau \bar{\bm{\Omega}}_n^{A,c} \right) - 2\left( g_n^{A,c} + \tau \bar g_n^{N,c} \right)\cdot \mathbf{E} \right].
\label{eq:B7}
\end{equation}

\twocolumngrid


\begin{thebibliography}{00}

\bibitem{Kane2010} %
M. Z. Hasan and C. L. Kane, Colloquium: Topological insulators,
\href{https://doi.org/10.1103/RevModPhys.82.3045}{Rev. Mod. Phys. \textbf{82}, 3045 (2010).}

\bibitem{SCZhang2011}
X.-L. Qi and S.-C. Zhang, Topological insulators and superconductors,
\href{https://doi.org/10.1103/RevModPhys.83.1057}{Rev. Mod. Phys. \textbf{83}, 1057 (2011).}

\bibitem{Das2016}
A. Bansil, Hsin Lin, and T. Das, Colloquium: Topological band theory,
\href{https://doi.org/10.1103/RevModPhys.88.021004}{Rev. Mod. Phys. \textbf{88}, 021004 (2016).}

\bibitem{Mermin1979}
N. D. Mermin, The topological theory of defects in ordered media,
\href{https://doi.org/10.1103/RevModPhys.51.591}{Rev. Mod. Phys. \textbf{51}, 591 (1979).}

\bibitem{Nakahara}
M. Nakahara, {\it{Geometry, Topology and Physics}} (2nd ed.),
\href{https://doi.org/10.1201/9781315275826}{(CRC Press, Boca Raton, 2003).}

\bibitem{Volovik2003}
G. E. Volovik, {\it{The Universe in a Helium Droplet}}, International Series of Monographs on Physics
\href{https://doi.org/10.1093/acprof:oso/9780199564842.001.0001}{(Oxford University Press, Oxford, 2003).}

\bibitem{Xiao2010}
D. Xiao, M.-C. Chang, and Q. Niu,
Berry phase effects on electronic properties,
\href{https://doi.org/10.1103/RevModPhys.82.1959}{Rev. Mod. Phys. \textbf{82}, 1959 (2010).}

\bibitem{AhnNP}   
J. Ahn, G.-Y. Guo, N. Nagaosa, and A. Vishwanath,
Riemannian geometry of resonant optical responses,
\href{https://doi.org/10.1038/s41567-021-01465-z}{Nat. Phys. \textbf{18}, 290 (2022).}

\bibitem{Onishi2024}
Y. Onishi and L. Fu,
Fundamental Bound on Topological Gap,
\href{https://doi.org/10.1103/PhysRevX.14.011052}{Phys. Rev. X \textbf{14}, 011052 (2024).}

\bibitem{Moore2010}
J. E. Moore and J. Orenstein,
Confinement-Induced Berry Phase and Helicity-Dependent Photocurrents,
\href{https://doi.org/10.1103/PhysRevLett.105.026805}{Phys. Rev. Lett. \textbf{105}, 026805 (2010).}

\bibitem{Ma2021}
Q. Ma, A. G. Grushin, and K. S. Burch,
Topology and geometry under the nonlinear electromagnetic spotlight,
\href{https://doi.org/10.1038/s41563-021-00992-7}{Nat. Mater. \textbf{20}, 1601 (2021).}

\bibitem{Liu2024}
T. Liu, X.-B. Qiang, H.-Z. Lu, X. C. Xie, Quantum geometry in condensed matter,
\href{https://doi.org/10.1093/nsr/nwae334}{Natl. Sci. Rev. \textbf{12}, nwae334 (2024).}

\bibitem{Komissarov2024}
I. Komissarov, T. Holder, and R. Queiroz,
The quantum geometric origin of capacitance in insulators,
\href{https://doi.org/10.1038/s41467-024-48808-x}{Nat. Commun. \textbf{15}, 4621 (2024).}

\bibitem{Xiang2024PRB}
L. Xiang, B. Wang, Y. Wei, Z. Qiao, and J. Wang,
Linear displacement current solely driven by the quantum metric,
\href{https://doi.org/10.1103/PhysRevB.109.115121}{Phys. Rev. B \textbf{109}, 115121 (2024).}

\bibitem{Tokura2018}
Y. Tokura and N. Nagaosa,
Nonreciprocal responses from non-centrosymmetric quantum materials,
\href{https://doi.org/10.1038/s41467-018-05759-4}{Nat. Commun. \textbf{9}, 3740 (2018).}

\bibitem{Holder2020}
T. Holder, D. Kaplan, and B. H. Yan,
Consequences of time-reversal-symmetry breaking in the light-matter interaction: Berry curvature, quantum metric, and diabatic motion,
\href{https://doi.org/10.1103/PhysRevResearch.2.033100}{Phys. Rev. Research \textbf{2}, 033100 (2020).}

\bibitem{Sodemann2015}
I. Sodemann and L. Fu,
Quantum Nonlinear Hall Effect Induced by Berry Curvature Dipole in Time-Reversal Invariant Materials,
\href{https://doi.org/10.1103/PhysRevLett.115.216806}{Phys. Rev. Lett. \textbf{115}, 216806 (2015).}

\bibitem{Ma2019}
Q. Ma, S.-Y. Xu, H. Shen, D. MacNeill, V. Fatemi, T.-R. Chang, A. M. M. Valdivia,
S. F. Wu, Z. Du, C.-H. Hsu, et al.,
Observation of the nonlinear Hall effect under time-reversal-symmetric conditions,
\href{https://doi.org/10.1038/s41586-018-0807-6}{Nature \textbf{565}, 337 (2019).}

\bibitem{Kang2019}
K. F. Kang, T. X. Li, E. Sohn, J. Shan, and K. F. Mak,
Nonlinear anomalous Hall effect in few-layer WTe$_2$,
\href{https://doi.org/10.1038/s41563-019-0294-7}{Nat. Mater. \textbf{18}, 324 (2019).}

\bibitem{Du2021}
Z. Z. Du, H.-Z. Lu, and X. C. Xie, Nonlinear Hall effects,
\href{https://doi.org/10.1038/s42254-021-00359-6}{Nat. Rev. Phys. \textbf{3}, 744 (2021).}

\bibitem{Lai2021}
S. Lai, H. Liu, Z. Zhang, J. Zhao, X. Feng, N. Wang, C. Tang, Y. Liu, K. S. Novoselov, S. Y. A. Yang, and W.-B. Gao,
Third-order nonlinear Hall effect induced by the Berry-connection polarizability tensor,
\href{https://doi.org/10.1038/s41565-021-00917-0}{Nat. Nanotechnol. \textbf{16}, 869 (2021).}

\bibitem{Wei2022PRB}
M. Wei, B. Wang, Y. Yu, F. Xu, and J. Wang,
Nonlinear Hall effect induced by internal Coulomb interaction and phase relaxation process in a four-terminal system with time-reversal symmetry,
\href{https://doi.org/10.1103/PhysRevB.105.115411}{Phys. Rev. B \textbf{105}, 115411 (2022).}

\bibitem{Wei2023PRL}
M. Wei, L. Wang, B. Wang, L. Xiang, F. Xu, B. Wang, and J. Wang,
Quantum Fluctuation of the Quantum Geometric Tensor and Its Manifestation as Intrinsic Hall Signatures in Time-Reversal Invariant Systems,
\href{https://doi.org/10.1103/PhysRevLett.130.036202}{Phys. Rev. Lett. \textbf{130}, 036202 (2023).}

\bibitem{Zhang2023}
C.-P. Zhang, X.-J. Gao, Y.-M. Xie, H. C. Po, and K. T. Law,
Higher-order nonlinear anomalous Hall effects induced by Berry curvature multipoles,
\href{https://doi.org/10.1103/PhysRevB.107.115142}{Phys. Rev. B \textbf{107}, 115142 (2023).}

\bibitem{Gao2023} %
A. Gao, Y.-F. Liu, J.-X. Qiu, B. Ghosh, T. V. Trevisan, Y. Onishi, C. Hu, T. Qian, H.-J. Tien, S.-W. Chen, \textit{et al.},
Quantum metric nonlinear Hall effect in a topological antiferromagnetic heterostructure,
\href{https://doi.org/10.1126/science.adf1506}{Science \textbf{381}, 181 (2023).}

\bibitem{Wang2023}
N.-Z. Wang, D. Kaplan, Z.-W. Zhang, T. Holder, N. Cao, A.-F. Wang, X.-Y. Zhou, F.-F. Zhou, Z.-Z. Jiang,
C. S. Zhang, et al.,
Quantum-metric-induced nonlinear transport in a topological antiferromagnet,
\href{https://doi.org/10.1038/s41586-023-06363-3}{Nature (London) \textbf{621}, 487 (2023).}

\bibitem{Kaplan2024}
D. Kaplan, T. Holder, and B.-H. Yan,
Unification of Nonlinear Anomalous Hall Effect and Nonreciprocal Magnetoresistance in Metals by the Quantum Geometry,
\href{https://doi.org/10.1103/PhysRevLett.132.026301}{Phys. Rev. Lett. \textbf{132}, 026301 (2024).}

\bibitem{Han2024}
J. Han, T. Uchimura, Y. Araki, J.-Y. Yoon, Y. Takeuchi, Y. Yamane, S. Kanai, J. Ieda, H. Ohno, and S. Fukami,
Room-temperature flexible manipulation of the quantum-metric structure in a topological chiral antiferromagnet,
\href{https://doi.org/10.1038/s41567-024-02476-2}{Nat. Phys. \textbf{20}, 1110 (2024).}

\bibitem{Jia2024}
J. X. Jia, L. J. Xiang, Z. H. Qiao, and J. Wang,
Equivalence of semiclassical and response theories for second-order nonlinear ac Hall effects,
\href{https://doi.org/10.1103/PhysRevB.110.245406}{Phys. Rev. B \textbf{110}, 245406 (2024).}

\bibitem{Wei2025PRL}
M. Wei, L. Xiang, F. Xu, B. Wang, and J. Wang,
Unconventional Hall Effect in Gapless Superconductors: Transverse Supercurrent Converted from Normal Current,
\href{https://doi.org/10.1103/fql8-f3tl}{Phys. Rev. Lett. \textbf{135}, 166304 (2025).}


\bibitem{XiangZeeman}
L. Xiang, J. Jia, F. Xu, Z. Qiao, and J. Wang,
Intrinsic Gyrotropic Magnetic Current from Zeeman Quantum Geometry,
\href{https://doi.org/10.1103/PhysRevLett.134.116301}{Phys. Rev. Lett. \textbf{134}, 116301 (2025).}

\bibitem{Zeeman1}
J. Cao, F. Qi, Y. Xiang, and G. Jin,
Magnetoelectric response arising from Zeeman quantum geometry,
\href{https://doi.org/10.1103/54qq-2qgh}{Phys. Rev. B \textbf{112}, 235145 (2025)}.

\bibitem{Zeeman2}
M. Ezawa,
Quantum geometry and X-wave magnets with X=p,d,f,g,i,
\href{https://doi.org/10.35848/1882-0786/ae4311}{Appl. Phys. Express \textbf{19}, 030101 (2026)}.

\bibitem{Zeeman3}
J. Tan, O. Matsyshyn, G. Vignale, and J. C. W. Song,
Nonequilibrium Exchange Nonlinear Hall Effect,
Preprint at \href{https://doi.org/10.48550/arXiv.2512.06074}{https://doi.org/10.48550/arXiv.2512.06074}.

\bibitem{Zeeman4}
N. Chakraborti, S. K. Ghosh, and S. Nandy,
Zeeman Quantum Geometry as a Probe of Unconventional Magnetism,
Preprint at \href{https://doi.org/10.48550/arXiv.2508.14745}{https://doi.org/10.48550/arXiv.2508.14745}.

\bibitem{Zhong2016}
S. Zhong, J. E. Moore, and I. Souza,
Gyrotropic Magnetic Effect and the Magnetic Moment on the Fermi Surface,
\href{https://doi.org/10.1103/PhysRevLett.116.077201}{Phys. Rev. Lett. \textbf{116}, 077201 (2016).}


\bibitem{Azizi2025}
A. Azizi, Hodge duals in spherical compactifications,
\href{https://doi.org/10.1103/57hd-7jrb}{Phys. Rev. D \textbf{112}, 026026 (2025).}

\bibitem{Gustafsson2022}
B. Gustafsson, Vortex pairs and dipoles on closed surfaces,
\href{https://doi.org/10.1007/s00332-022-09822-9}{J. Nonlinear Sci. \textbf{32}, 62 (2022).}


\bibitem{Montambaux2018}
G. Montambaux, L.-K. Lim, J.-N. Fuchs, and F. Pi\'{e}chon,
Winding Vector: How to Annihilate Two Dirac Points with the Same Charge,
\href{https://doi.org/10.1103/PhysRevLett.121.256402}{Phys. Rev. Lett. \textbf{121}, 256402 (2018).}

\bibitem{Furukawa2017}
T. Furukawa, Y. Shimokawa, K. Kobayashi, and T. Itou,
Observation of current-induced bulk magnetization in elemental tellurium,
\href{https://doi.org/10.1038/s41467-017-01093-3}{Nat. Commun. \textbf{8}, 954 (2017).}

\bibitem{Xu2020}
B. Xu, Z. Fang, M.-\'{A}. S\'{a}nchez-Mart\'{\i}nez, J. W. F. Venderbos, Z. Ni, T. Qiu, K. Manna, K. Wang, J. Paglione, C. Bernhard, C. Felser, E. J. Mele, A. G. Grushin, A. M. Rappe, and L. Wu,
Optical signatures of multifold fermions in the chiral topological semimetal CoSi,
\href{https://doi.org/10.1073/pnas.2010752117}{PNAS \textbf{117}, 27104 (2020).}

\bibitem{Ni2021}
Z. Ni, K. Wang, Y. Zhang, O. Pozo, B. Xu, X. Han, K. Manna, J. Paglione, C. Felser, A. G. Grushin, F. de Juan, E. J. Mele, and L. Wu,
Giant topological longitudinal circular photo-galvanic effect in the chiral multifold semimetal CoSi,
\href{https://doi.org/10.1038/s41467-020-20408-5}{Nat. Commun. \textbf{12}, 154 (2021).}

\bibitem{Xiang-spin}
L. Xiang, H. Jin, and J. Wang,
Spin Transport Revealed by Spin Quantum Geometry,
\href{https://doi.org/10.1103/14mp-263q}{Phys. Rev. Lett. \textbf{135}, 146303 (2025).}

\bibitem{universal}
J. Sinova, D. Culcer, Q. Niu, N. A. Sinitsyn, T. Jungwirth, and A. H. MacDonald,
Universal Intrinsic Spin Hall Effect,
\href{https://doi.org/10.1103/PhysRevLett.92.126603}{Phys. Rev. Lett. \textbf{92}, 126603 (2004).}

\bibitem{Mertig}
A. Mook, R. R. Neumann, A. Johansson, J. Henk, and I. Mertig,
Origin of the magnetic spin Hall effect: Spin current vorticity in the Fermi sea,
\href{https://doi.org/10.1103/PhysRevResearch.2.023065}{Phys. Rev. Research \textbf{2}, 023065 (2020)}.

\bibitem{W-Yao}
D. Zhai, C. Chen, C. Xiao, and W. Yao,
Time-reversal even charge Hall effect from twisted interface coupling,
\href{https://doi.org/10.1038/s41467-023-37644-0} {Nat. Commun. \textbf{14}, 1961 (2023).}

%
%

%


\bibitem{Aversa1995}
C. Aversa and J. E. Sipe,
Nonlinear optical susceptibilities of semiconductors: Results with a length-gauge analysis,
\href{https://doi.org/10.1103/PhysRevB.52.14636}{Phys. Rev. B
\textbf{52}, 14636 (1995).}
\end{thebibliography}
\end{document}